\documentclass[journal]{IEEEtran}
\usepackage{xcolor,soul,framed} %,caption
 
\colorlet{shadecolor}{yellow}
\usepackage[pdftex]{graphicx}
\graphicspath{{../pdf/}{../jpeg/}}
\DeclareGraphicsExtensions{.pdf,.jpeg,.png}

\usepackage[cmex10]{amsmath}
\usepackage{array}
\usepackage{float}
\usepackage{booktabs} % For professional-looking tables
\usepackage{multirow}
\usepackage{mdwmath}
\usepackage{mdwtab}
\usepackage{graphicx}
\usepackage{eqparbox}
\usepackage{url}
\usepackage{amsmath}
\usepackage{amssymb}
\usepackage[hyperindex, hidelinks]{hyperref} 
\hypersetup{
    colorlinks=true,       % 启用颜色链接
    linkcolor=blue,        % 设置链接颜色为蓝色
    citecolor=blue,        % 设置文献引用的颜色为蓝色
    urlcolor=blue          % 设置网址链接的颜色为蓝色
}
\usepackage{subcaption}
\usepackage{algorithm}
\usepackage{algpseudocode}
\begin{document}
\bstctlcite{IEEEexample:BSTcontrol}
    \title{BrownoutServe: SLO-Aware Inference Serving under Bursty Workloads for MoE-based LLMs}
  \author{Jianmin~Hu,
      Minxian~Xu,~\IEEEmembership{Senior~Member,~IEEE,}\\
      Kejiang~Ye,~\IEEEmembership{Senior~Member,~IEEE,}
      and~Chengzhong~Xu,~\IEEEmembership{Fellow,~IEEE}% <-this % stops a space

  \thanks{This work is supported by Guangdong Basic and Applied Basic Research Foundation (No. 2024A1515010251, 2023B1515130002), Guangdong Special Support Plan (No. 2021TQ06X990), Shenzhen Basic Research Program under grants JCYJ20220818101610023, and JCYJ20240809180935001, Shenzhen Industrial Application Projects of undertaking the National key R \& D Program of China (No. CJGJZD20210408091600002). 
  
  Jiamin Hu is with Shenzhen Institutes of Advanced Technology, Chinese
  Academy of Sciences, and Southern University of Science and Technology, Shenzhen 518055.
  
  Minxian Xu and Kejiang Ye are with the Shenzhen Institutes of Advanced
  Technology, Chinese Academy of Sciences, Shenzhen 518055, China.
  
  Chengzhong Xu is with the State Key Lab of IOTSC, University of Macau,
  Macau 999078, China.

  Minxian Xu is the corresponding author (e-mail:
  mx.xu@siat.ac.cn).
  }
}

% \markboth{IEEE TRANSACTIONS ON PARALLEL AND DISTRIBUTED SYSTEMS}

% ====================================================================
\maketitle

% === ABSTRACT ====================================================================
% =================================================================================
\begin{abstract}
%\boldmath
In recent years, the Mixture-of-Experts (MoE) architecture has been widely applied to large language models (LLMs), providing a promising solution that activates only a subset of the model's parameters during computation, thereby reducing overall memory requirements and allowing for faster inference compared to dense models. Despite these advantages, existing systems still face issues of low efficiency due to static model placement and lack of dynamic workloads adaptation. This leads to suboptimal resource utilization and increased latency, especially during bursty requests periods.

To address these challenges, this paper introduces BrownoutServe, a novel serving framework designed to optimize inference efficiency and maintain service reliability for MoE-based LLMs under dynamic computational demands and traffic conditions. BrownoutServe introduces “united experts" that integrate knowledge from multiple experts, reducing the times of expert access and inference latency. Additionally, it proposes a dynamic brownout mechanism to adaptively adjust the processing of certain tokens, optimizing inference performance while guaranteeing service level objectives (SLOs) are met. Our evaluations show the effectiveness of BrownoutServe under various workloads: it achieves up to 2.07× throughput improvement compared to vLLM and reduces SLO violations by 90.28\%, showcasing its robustness under bursty traffic while maintaining acceptable inference accuracy.
\end{abstract}

% === KEYWORDS ====================================================================
% =================================================================================
\begin{IEEEkeywords}
MoE, LLM inference serving, SLO, brownout, requests burst
\end{IEEEkeywords}

\section{Introduction}
\IEEEPARstart{M}{oE}~\cite{jacobs1991adaptive} architecture has garnered widespread attention in the field of LLMs, with a surge of research and applications in recent years, such as Mixtral~\cite{jiang2024mixtral}, Qwen3~\cite{yang2025qwen3}, DeepSeekV3~\cite{liu2024deepseek}, and their variants. MoE architectures significantly reduce training and inference costs by activating only a subset of model parameters (experts) during computation~\cite{lewis2021base}. However, efficient inference for MoE models remains a challenging problem in both industry and academia.

\begin{figure}[tbp]
  \centering
  \begin{minipage}[b]{0.24\textwidth}
    \centering
    \includegraphics[width=\textwidth]{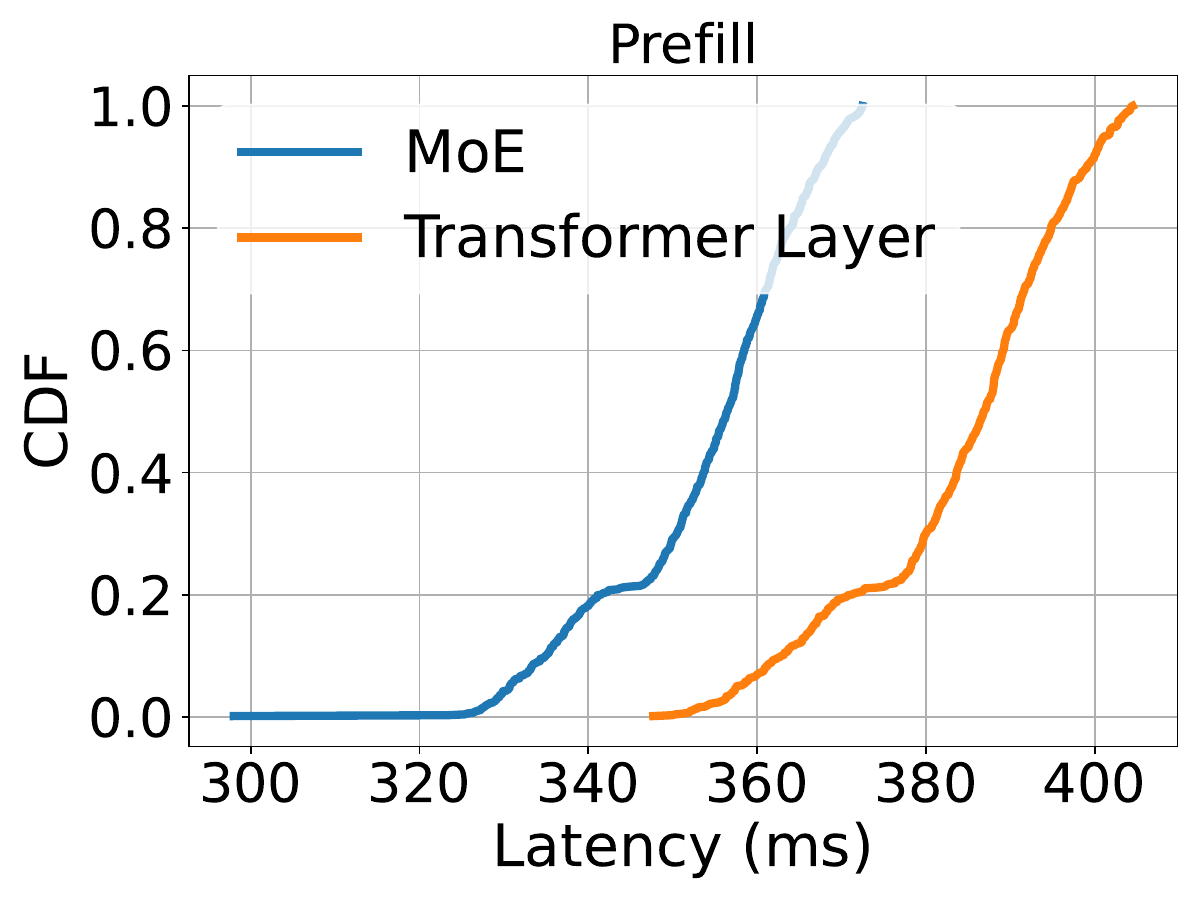}

    \label{fig:prefill_cdf}
  \end{minipage} \hfill
  \begin{minipage}[b]{0.24\textwidth}
    \centering
    \includegraphics[width=\textwidth]{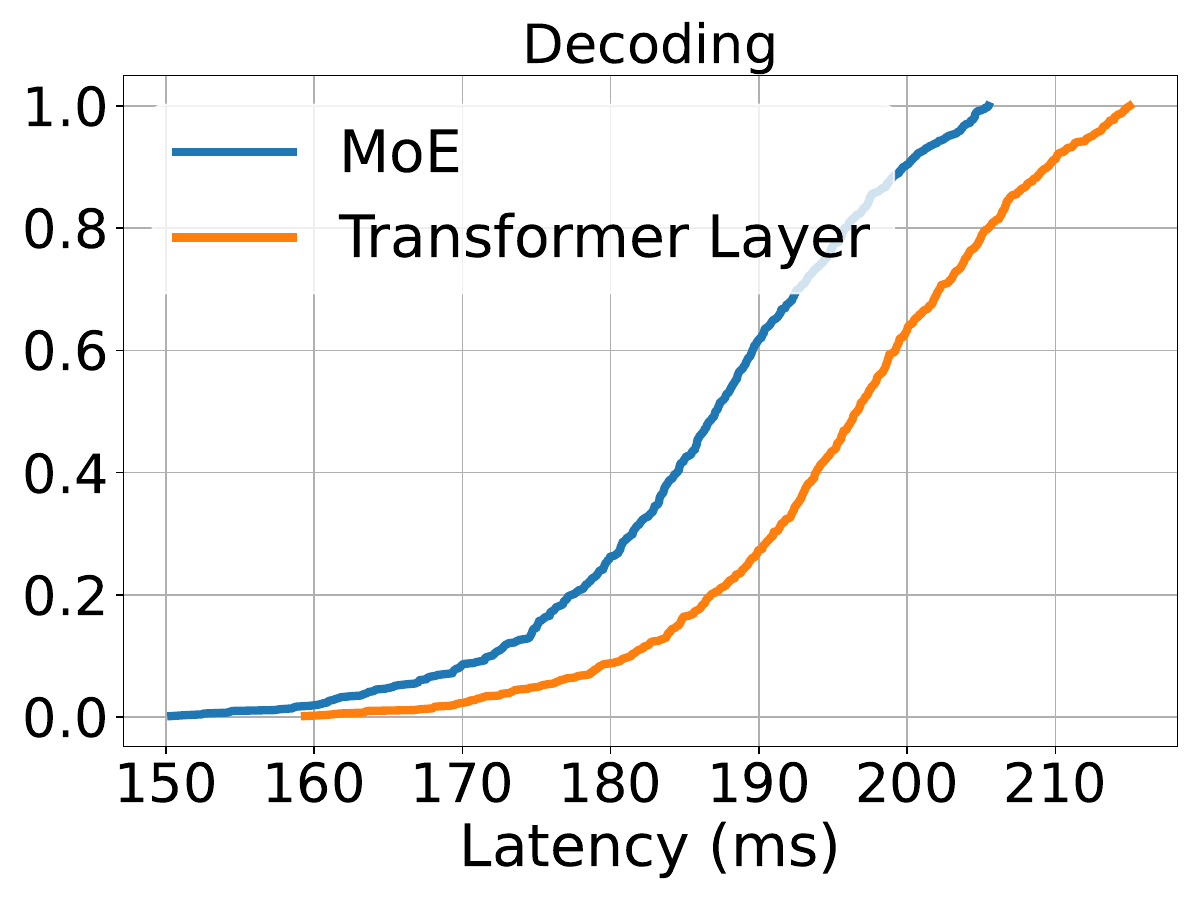}

    \label{fig:decoding_cdf}
  \end{minipage}
  \caption{Cumulative distribution function (CDF) of latency for the prefill and decoding stages of the MoE module and transformer layer using the Qwen1.5-MoE-A2.7B model with a batch size of 64 on the Alpaca dataset, utilizing four A100-40GB-PCIE GPUs. The left plot illustrates the latency distribution for the prefill stage, and the right plot shows the latency distribution for the decoding stage.}
    \label{fig:cdf}
\end{figure}

LLM inference is divided into two stages: prefill and decoding~\cite{zhong2024distserve}. During the prefill stage, the model processes the user's input to generate the first output token, caching the generated key-value pairs in the key-value (KV) cache to avoid redundant computation. This stage involves processing a large number of tokens, activating almost all experts, and often becomes a performance bottleneck. In the decoding stage, each request processes only one token at a time, but when the model is under high load (i.e., the current batch size approaches the maximum batch size), the MoE module can still become a bottleneck. Studies~
\cite{lewis2021base,wolf2020transformers,fedus2022switch} have shown that due to the varying activity levels of different experts, only a few experts handle a large number of tokens (hot experts), while most experts handle very few tokens (cold experts). This imbalance prevents these cold experts from fully utilizing GPU parallelism, leading to decreased resource utilization. Fig.~\ref{fig:cdf} illustrates that in the prefill stage, the latency of the MoE accounts for about 81.23\% of the total latency of the transformer layer (including attention and MoE), and in the decoding stage, the MoE latency accounts for about 93.89\% of the total latency of the transformer layer. Therefore, reducing the latency of the MoE module is crucial to reducing the overall inference latency.

Existing MoE inference systems, such as vLLM~\cite{kwon2023efficient} (an engine optimized for both MoE and dense models) and DeepSpeed-MoE~\cite{rajbhandari2022deepspeed} (an engine specifically optimized for MoE models), have introduced techniques like tensor parallelism and expert parallelism to accelerate inference. However, these solutions are mostly statically configured and fail to handle sudden load spikes and request fluctuations, often leading to SLO violations. In low-resource GPU clusters, such as those with a limited number of GPUs, restricted memory capacity, and constrained communication bandwidth, MoE inference faces several key challenges: uneven expert load causing computational bottlenecks, high cross-GPU communication costs becoming the performance floor if using parallelism techniques, and large model size making it costly and inflexible to scale up in response to bursty requests. These challenges significantly limit the usability and practicality of MoE models in small-to-medium-scale deployment environments with limited resources.

To address these challenges, we introduce the concept of united experts and brownout approach to reduce inference latency of MoE module. United experts integrate the knowledge of multiple experts into a single expert of equivalent parameter size, reducing the times of expert access during inference, increasing the average number of tokens processed by each expert, and fully leveraging GPU parallelism to reduce latency and improve throughput. The brownout approach is inspired by the brownout strategies in power systems. This approach has also been applied in cloud computing environment that can dynamically activate or deactivate optional application components based on different workloads~\cite{xu2020self}. In this work, the brownout approach is used to determine which tokens are processed by the original experts and which are processed by the united experts with accuracy and efficiency trade-offs. To mitigate this, based on the brownout approach, we propose the SLO-Aware Latency Control algorithm, which dynamically adjust the brownout configuration to minimize accuracy loss while ensuring inference latency meets SLO, even under fluctuating request rate and bursty workloads.

In this paper, we present BrownoutServe, an efficient MoE inference serving framework designed for small-to-medium-scale GPU clusters. The core design of BrownoutServe includes two key mechanisms: First, the united expert model, which integrates the knowledge of multiple experts into a united expert to reduce expert access during inference. Second, the dynamic brownout mechanism (brownout approach and SLO-Aware Latency Control algorithm), which dynamically routes a subset of tokens to the united experts under resource constraints or bursty workloads, reducing expert access overhead and helping maintain SLO attainment. Our evaluations demonstrate effectiveness of BrownoutServe across various workloads. Our contributions are as follows:
\begin{itemize}
\item We develop an efficient inference framework for MoE-based LLMs, featuring a united experts-based MoE module and an SLO-Aware Latency Control algorithm-driven scheduler.
\item We introduce the concept of united experts and the brownout approach, both of which can be flexibly adapted in scenarios with bursty workloads, and elaborate on three brownout strategies: zero-brownout, full-brownout, and partial-brownout.
\item We provide concise implementations of the brownout approach and the SLO-Aware Latency Control algorithm.
\item We comprehensively evaluate BrownoutServe's performance across various realistic workloads.  Compared to vLLM, BrownoutServe achieves up to 2.07× higher throughput and reduces SLO violations by 90.28\%.
\end{itemize}
\section{Related Work}
In this section, we discuss the related work on inference optimization for LLMs. Current work can be categorized into two classes as follows:

\textbf{Optimization for MoE-based LLMs.}
As MoE-based architectures evolve and model parameters expand exponentially, a multitude of optimization approaches have emerged. Representative work includes GShard~\cite{lepikhin2020gshard} which significantly improves computational efficiency and scalability by distributing expert modules of the MoE layer across multiple devices. Switch Transformer~\cite{fedus2022switch} simplified routing by activating only one expert per token. Lina~\cite{li2023accelerating} uses dynamic resource scheduling and expert popularity estimation to balance the workload across devices and improve inference efficiency. MPMoE~\cite{zhang2024mpmoe} introduce adaptive pipeline parallelism and memory‑reuse strategies to accelerate training and reduce memory overhead. EfficientMoE~\cite{zeng2025efficientmoe} uses dynamic scheduling and capacity adjustment to optimize the training efficiency of MoE models. Tutel~\cite{hwang2023tutel} presents a highly scalable stack design and implementation for MoE, featuring dynamically adaptive parallelism and pipelining. Pre-gated MoE~\cite{hwang2024pre} introduces a novel pre-selection mechanism (pre-gating function), which predicts the set of experts to be activated before inference. This allows for pre-loading necessary parameters, thereby reducing the overhead of dynamic scheduling during inference. MoE-Lightning~\cite{cao2025moe} implements CPU-GPU-I/O pipeline scheduling and a Hierarchical Roofline Model, enabling high-throughput MoE inference on memory-constrained GPUs. Although these optimization techniques have accelerated inference to some extent, they lack the ability to handle bursty requests, thereby causing SLO violations. Our work addresses these limitations by introducing novel mechanisms such as united experts and the brownout approach to dynamically optimize inference performance and maintain service reliability under fluctuating workloads.

\textbf{Optimization for generic LLMs}
There are many related works focused on optimizing inference for generic LLMs. For KV cache optimization, vLLM~\cite{kwon2023efficient} proposes PagedAttetion for KV cache management, which reduces GPU memory fragmentation. InfiniGen~\cite{lee2024infinigen} reduces data transfer overhead by dynamically selecting and prefetching only the essential KV cache entries instead of loading all of them. For parallelism strategies optimization, LoongServe~\cite{wu2024loongserve} provides an Elastic Sequence Parallelism strategy, which effectively improves resource utilization when serving long-context LLMs. alpaServe~\cite{li2023alpaserve} leverages model parallelism: even when a model fits on a single device, it can be partitioned across multiple devices to enable statistical multiplexing, allowing better handling of bursty requests. DistServe~\cite{zhong2024distserve} decouples the prefill and decoding phases to eliminate interference between them and applies separate parallelism strategies for each stage. For scheduling optimization, Orca~\cite{yu2022orca} proposes iteration-level scheduling, a more fine-grained scheduling approach where completed requests are removed from the batch after each iteration, and new requests can be added to the batch for the next iteration. Luminix~\cite{sun2024llumnix} introduces a dynamic scheduling mechanism inspired by context switching in operating systems, which enables real-time reallocation of requests among multiple model instances, achieving better load balancing and resource isolation. These optimization techniques, while originally designed for generic LLMs, are still applicable to MoE-based LLMs. BrownoutServe is orthogonal to these techniques and can work in conjunction with them. Therefore, BrownoutServe integrates some of these techniques, such as using PagedAttention to optimize KV cache management and adopting iteration-level scheduling to improve the scheduler performance. These integrations help BrownoutServe achieve better overall optimization for MoE-based LLMs.
\section{Background and Motivation}

In this section, we briefly introduce MoE-based LLMs and the theoretical foundations of service. We then reveal the two main challenges currently faced by MoE-based LLMs in real-world deployment.

\subsection{Mixture-of-Experts}
The MoE-based LLMs replace the feed-forward networks (FFNs) in standard Transformer-based~\cite{wolf2020transformers} LLMs with a gating function and multiple small-parameter FFNs, enabling sparse activation and improved computational efficiency. Compared to traditional dense models~\cite{zhang2022opt,touvron2023llama,achiam2023gpt}, MoE offers a solution to reduce computational and memory requirements by activating only a subset of the model's parameters, known as ``experts'', during computation. This selective activation mechanism is able to reduces the overall memory footprint and computational load, enabling more efficient use of resources and facilitating the training and deployment of models with trillions of parameters.
Like dense models, the inference process for MoE typically involves two stages: the prefill stage, where the model generates an initial output token based on the user's input, and the decoding stage, where subsequent tokens are generated one at a time. The MoE architecture is particularly beneficial during these stages as it can adjust the number of active experts based on the input data, leading to more efficient processing and reduced latency compared to dense models. Each expert represents a small portion of the knowledge contained within the model, thus for each incoming token, it first needs to undergo computation by a gating function to determine which experts should process it next.

\subsection{Theoretical Foundations of LLM Service Modeling}
\textbf{Request Modeling:} In LLM services, user requests arrive randomly and can be modeled as a Poisson process, where the inter-arrival times follow an exponential distribution. This assumption is widely adopted in network and distributed system analysis. Given that the inference time per request is nearly constant (especially when the batch size is 1), the system can be modeled as a classic $M/D/1$ queue~\cite{kendall1953stochastic}, and the total average response time $W$ can be approximately expressed as:
\begin{equation}
W = \frac{\lambda \tau^2}{2(1 - \lambda \tau)} + \tau \label{eq:request_modeling},
\end{equation}
where $\lambda$ is the average request arrival rate, and $\tau$ is the fixed service time per request. This expression captures both the waiting time (the first term) in the queue and the actual processing time (the second term), and it highlights that as the system approaches saturation (i.e., $\lambda\tau \rightarrow 1$) the response time grows rapidly.

\textbf{System Optimization:}
According to Amdahl’s Law~\cite{amdahl1967validity}, the overall system speedup is limited by the proportion of the workload that is not optimized. The formula is as follows: 
\begin{equation}
Speedup = \frac{1}{(1-\alpha)+\frac{\alpha}{K}} \label{eq:speedup},
\end{equation}
where $\alpha$ represents the fraction of execution time consumed by the bottleneck component in the original system, $K$ denotes the speedup factor achieved by optimizing that component. This law emphasizes that the overall performance improvement of a system depends on the extent to which the performance bottleneck is optimized. For example, suppose the bottleneck component accounts for $\alpha = 0.6$ (i.e., 60\%) of the total processing time. If we optimize it by a factor of $K=3$ through parallelization, kernel fusion, or communication optimization, the overall system speedup would be:

\begin{equation}
Speedup = \frac{1}{(1 - 0.6) + \frac{0.6}{3}} = \frac{1}{0.4 + 0.2} = \frac{1}{0.6} \approx 1.67.
\end{equation}
This result illustrates that focusing on a single bottleneck can substantially  enhance system performance.
% \fi

% ====================================================================
% ====================================================================
% ====================================================================

\subsection{Challenges in MoE-base LLMs Inference Serving}

Although the MoE architecture significantly reduces computation and memory costs during the training and inference stages of LLMs, it also introduces some systemic challenges during real-world deployment and online serving. These challenges mainly fall into the following two categories:

\textbf{C1: Computational Pressure with Large Batch Size and Experts Load Imbalance.}
Compared to dense models, MoE models exhibit a much larger total number of parameters due to the presence of multiple experts, even though each inference path only activates a sparse subset of them. In scenarios with large batch sizes, a high number of tokens are routed to experts simultaneously, potentially activating nearly all experts. This turns the inference process from sparse to quasi-dense, resulting in intense compute and memory bandwidth pressure on the system.
Although top-k routing strategies (e.g., Top-2) used in GShard~\cite{lepikhin2020gshard} and Switch Transformer~\cite{fedus2022switch} aim to limit the number of experts each token activates, higher values of k are often required in practice to maintain model accuracy. This leads to frequent activation of most experts, especially under large batches, causing communication bandwidth bottlenecks in multi-GPU or multi-node deployments.
Moreover, token distribution across experts typically exhibits a long-tail pattern: only a small portion of the experts handle more tokens than average, while the majority are severely underutilized. This load imbalance results in idle compute resources and exacerbates performance degradation due to communication overhead and synchronization latency, particularly in distributed setups.  

\textbf{C2: SLO Violations under Bursty Workloads.}  Bursty workloads is a common scenario across various service types, and MoE-based serving systems struggle to meet SLOs under such conditions, especially in GPU clusters with limited resources. Most mainstream elasticity mechanisms still rely on traditional horizontal scaling methods, such as AWS EC2 instance spawning or Kubernetes~\cite{kubernetes2025} HPA-based auto-scaling, to handle bursty workloads. However, these approaches suffer from high cold start latency and substantial GPU memory overhead. Cloud instances typically take 30–60 seconds to initialize. MoE models often require 30–90 seconds to load weights due to their large size (tens of GBs);   Additional delays stem from container cold starts, network topology reconstruction, requests forwarding. These latencies can lead to total startup latencies exceeding 1–2 minutes, which is unacceptable for latency-sensitive applications such as chatbot or real-time recommendation systems. This untimely scaling often leads to substantial SLO violations. Although emerging systems such as ServerlessLLM~\cite{fu2024serverlessllm} attempt to reduce cold start latency through mechanisms like lazy weight loading and model migration, these approaches often rely on high-speed local caching and high-bandwidth environments, placing greater pressure on GPU memory capacity and network I/O. Additionally, horizontally scaling LLM instances often requires provisioning multiple full copies of the model, leading to memory consumption that is several times higher than the unscaled baseline~\cite{shubha2024usher}. This results in considerably increased hardware costs for cloud providers. What exacerbates the problem is that bursty traffic is often short-lived. Once the workloads subsides, the provisioned GPU instances remain underutilized or idle, incurring inefficient resource usage and wasted infrastructure cost. Such elasticity mismatches highlight the limitations of traditional scaling approaches in handling highly dynamic LLM workloads.

To tackle C1, we propose the concept of the united experts, which merges multiple experts into a single, consolidated expert. This can reduce the number of experts involved in computation, thereby alleviating the computational pressure during inference. Additionally, by combining underutilized experts that receive relatively few tokens, we can mitigate the issue of low GPU utilization, leading to more efficient resource usage and better system throughput.

To tackle C2, we introduce the brownout approach along with the SLO-Aware Latency Control (SALC) algorithm to reduce latency. This mechanism dynamically adjust a part of tokens processed by united experts based on real-time inference latency. This adaptive token routing helps reduce the overall latency introduced by the MoE module, which is often the bottleneck in the inference, while maintaining acceptable accuracy levels. As shown in Eq.~(\ref{eq:speedup}), since the MoE module dominates end-to-end latency in typical LLM inference workloads, applying the brownout approach and SALC algorithm effectively lowers total system latency, enabling the system to better meet SLO requirements under bursty traffic conditions.
\section{Design and Implementation of BrownoutServe}\label{DI}
In this section, we present BrownoutServe, an efficient and reliable inference framework for MoE-based LLMs, specifically designed to handle bursty workloads while maintaining SLO attainment. 
\subsection{Overview}\label{sec:overview}
\begin{figure}[tbp]
  \begin{center}
  \includegraphics[width=3.5in]{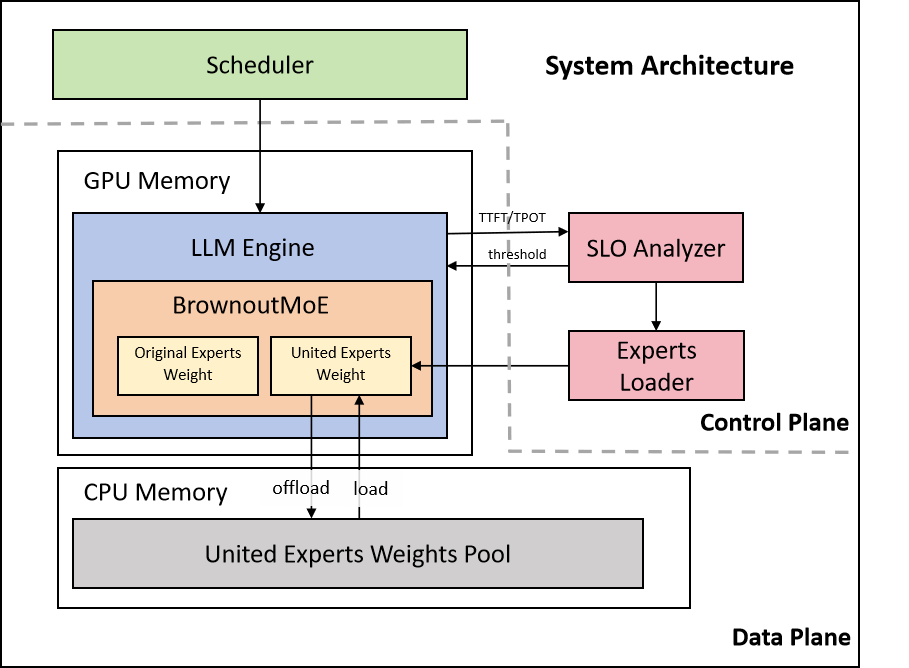}\\
\caption{Overview of BrownoutServe design.}
  \label{fig:}
  \end{center}
\end{figure}

BrownoutServe is an end-to-end MoE-based serving framework and its overview is shown in Fig.~\ref{fig:}. BrownoutServe adopts a modular design that divides the entire system into two major parts: the control plane and the data plane. The control plane consists of \texttt{Scheduler}, \texttt{SLO Analyzer}, and \texttt{Experts Loader}, while the data plane is comprised of a highly optimized LLM inference engine that integrates the BrowoutMoE module mentioned in section~\ref{sec:BrownoutMoE}. Additionally, the BrowoutMoE module includes a pluggable fused MoE~\cite{gale2023megablocks}. The LLM engine also incorporates existing optimization techniques, such as FlashAttention~\cite{dao2022flashattention}, PagedAttention~\cite{kwon2023efficient}, and ContinuousBatching~\cite{yu2022orca}. The entire system is implemented using approximately 5.5k lines of Python code.

\textbf{Control Plane.} \texttt{Scheduler} uses the first-come first-served (FCFS) algorithm to schedule incoming requests. When the engine reaches its maximum batch processing capacity, excess requests are stored in a waiting queue and processed in sequence. The scheduler supports streaming inputs and outputs, allowing for the dynamic insertion of new requests and the early removal of completed requests. \texttt{SLO Analyzer} continuously monitors the LLM engine's inference metrics, including time-to-first-token (TTFT) and time-per-output-token (TPOT), and uses the SLO-Aware Latency Control algorithm to timely adjust the brownout configuration to reduce SLO violations. \texttt{Experts Loader} is responsible for loading and unloading the united experts, as well as updating the way of united experts.

\textbf{Data Plane.} The LLM inference engine is primarily built based on PyTorch~\cite{paszke2019pytorch}. For some operations, such as PagedAttention and FlashAttention, they are implemented using Triton~\cite{tillet2019triton} language instead of traditional C++/CUDA, because of Triton's seamless compatibility with PyTorch. Compared to PagedAttention proposed in vLLM,  we optimize it by moving the block table to the GPU and implement block table operations as kernels (functions executed on the GPU), fully leveraging the parallel computing capability of the GPU to effectively reduce the additional overhead incurred by PagedAttention. The MoE module introduces the brownout approach, and to further optimize performance, we have also rewritten the MoE-related operators using Triton.

\subsection{United Expert Model}\label{sec:united_expert_model}
We introduce a novel model in MoE architecture, termed the “united expert". This is an efficient expert model by integrating the knowledge from multiple expert models. This innovative approach alleviates the issue of insufficient GPU computational resources and under-utilization in some unpopular experts, reduces the frequency of expert access, thereby effectively decreasing inference latency and enhancing the efficiency of model inference services.

For each transformer layer, which contains $m$ experts (subsequently called original experts), we divide these $m$ experts into 
 $\lceil \frac{m}{k} \rceil$  groups, where $k$ is the way of united experts (the number of experts in each group). For each group of original experts, we train a united expert via knowledge distillation~\cite{hinton2015distilling}, enabling it to take over their work within the brownout mechanism.
During the training process, all original experts in this group serve as the teacher model~\cite{hinton2015distilling} and this united expert is used as the student model~\cite{hinton2015distilling}. The output of the original experts (i.e. hidden states) will be used as the training target for the united experts. In this process, by minimizing the difference between the hidden states of the united expert and original experts in this group, the united expert model will learn the comprehensive knowledge of the original experts. Through this method, the united expert model can learn to approximate the knowledge representation of each group of original experts, thereby reducing the times of model access and maintaining the model's inference accuracy across various tasks. We use the mean squared error (MSE) loss function to measure the difference between the united expert and the original experts, which can be mathematically described as follows:
\begin{equation}
\mathcal{L}^{j}_{\mathrm{MSE}} = \frac{1}{k} \left( \sum_{i=0}^{k-1} \left\| H^{j}_{u} - H^{j \times k + i}_{o} \right\|^{2} \right), 
\end{equation}
where $\mathcal{L}_{\text{MSE}}^{j}$ is the difference between the original experts and the united expert for the j-th group. $k$ is the number of original experts in each group. $H_{\text{u}}^j$ is the hidden states of the j-th united expert, and $H_{\text{o}}^{j\times k+i}$ is the hidden states output by the i-th original expert in the j-th group, since it is grouped according to the expert index, $H_{\text{o}}^{j\times k+i}$ can also be understood as the ($j\times k+i$)-th original expert of the transformer layer.  
 The range of $j$ is from 0 to $\lceil \frac{m}{k} \rceil - 1 $. After training, we obtain $\lceil \frac{m}{k} \rceil$ united experts in each transformer layer.

\begin{figure}[tbp]
  \begin{center}
  \includegraphics[width=3.5in]{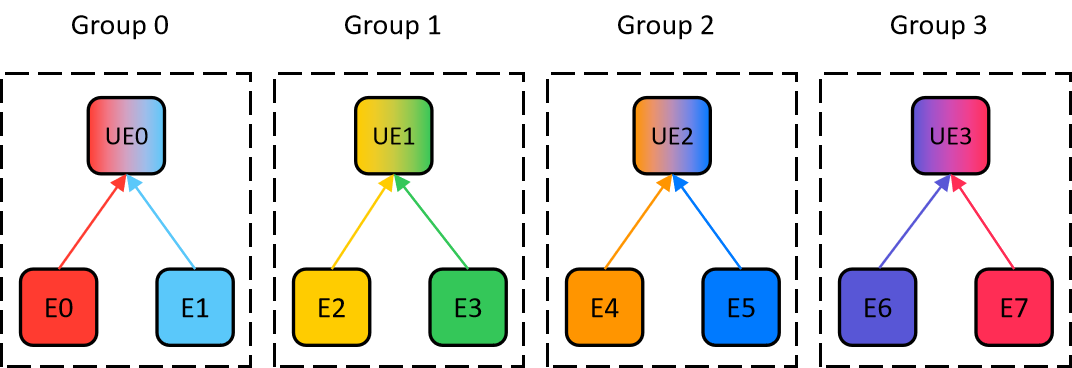}\\
\caption{Training process of four united experts (UE0-UE3), each integrating knowledge from eight original experts (E0-E7), color-coded to represent different token subsets.}
  \label{fig:united_experts}
  \end{center}
    % \vspace{-1.2em} 
\end{figure}

Fig.~\ref{fig:united_experts} shows a specific example of training process for united experts. In this example, we have 8 original experts and divide every 2 experts into groups, resulting in $\lceil \frac {8} {2} \rceil=4$ groups. We can use the method described above to train a united expert for each group, which generally includes the knowledge of all original experts in the group. For instance, for Group 0, in a subsequent inference iteration, if the number of tokens processed by E0 and E1 is relatively small, we can concatenate the tokens to be processed by these two original experts together and provide them to the united expert UE0 of the group for processing, in order to improve GPU computing utilization.

% =======
% FIG. 01
% =======

\subsection{Brownout Approach} \label{sec:brownout_approach}
The application of the brownout approach in MoE-based serving  draws an analogy from the brownout mechanisms of the power system. In scenarios where power resources are strained or face a surge in demand, power companies may implement brownout strategies, temporarily disconnecting non-essential facilities to guarantee the power supply to critical infrastructure. Similarly, we propose the brownout approach within the MoE architecture. When computational resources are limited or there is a sudden bursty of user requests, the brownout strategy can be employed to degrade the processing of certain tokens, thereby optimizing the model's inference efficiency and guaranteeing SLOs. In this subsection, we introduce three brownout strategies: zero-brownout, full-brownout, and partial-brownout.

\begin{figure}[tbp]
  \centering
  \begin{minipage}{0.24\textwidth}
    \centering
    \includegraphics[width=\textwidth]{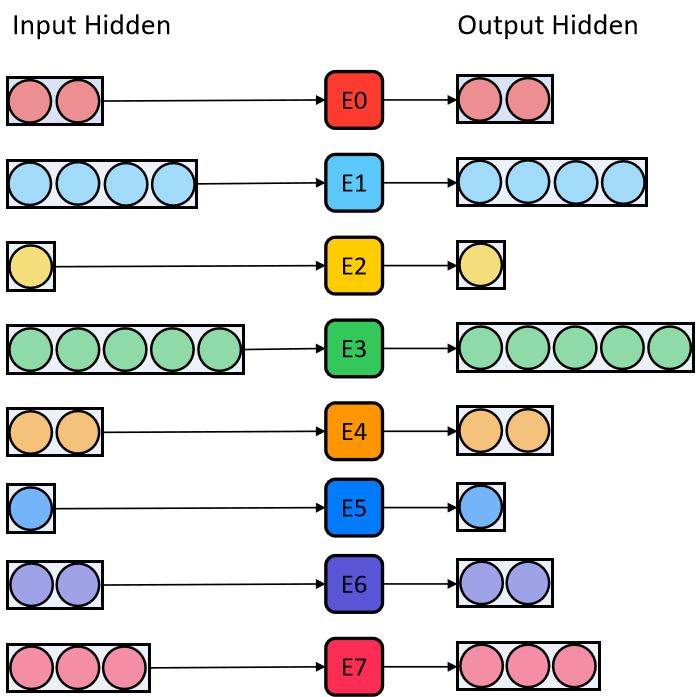}
    \subcaption{Zero-brownout}
    \label{fig:Zero-brownout}
  \end{minipage} \hfill
  \begin{minipage}{0.24\textwidth}
    \centering
    \includegraphics[width=\textwidth]{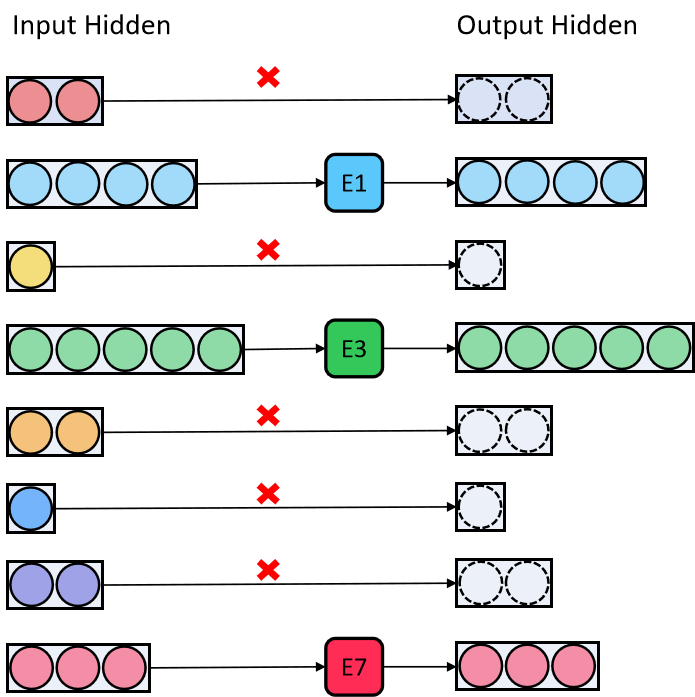}
    \subcaption{Full-brownout}
    \label{fig:Full-brownout}
  \end{minipage}
  \caption{Two brownout strategies illustration with 8 experts (E0-E7) processing 20 tokens.}
    \label{fig:two brownout strategies}
    \vspace{-1em}
\end{figure}

\textbf{Zero-Brownout.} Zero-brownout is a strategy that does not implement any degrading measures, meaning that the number of expert model activations is never reduced under any circumstances. Under this strategy, each input token is processed by all relevant original expert models. This strategy ensures that all tokens are processed by original experts, thereby providing the highest theoretical accuracy in model inference. Existing MoE inference engines can be considered as adopting zero-brownout, suitable for environments with ample computational resources and stable user request loads. Fig.~\ref{fig:Zero-brownout} shows a simple example of zero-brownout. There are a total of 8 experts and 20 tokens. After passing through the gate unit, the tokens are allocated to the corresponding experts for processing, and no tokens are degraded or dropped out.

\textbf{Full-Brownout.} The idea of full-brownout lies in strictly controlling the number of expert activations in the model. Considering that expert access is the bottleneck of the entire model's inference performance, reducing the times of expert access appropriately can effectively reduce the inference latency to meet user SLO requirements. This mechanism requires setting a $threshold \ (0\leq threshold \leq 1)$ in advance, which determines the proportion of tokens that need to be processed by original experts, while other tokens will be ignored in current transformer layer. To minimize inference latency, we need to find fewer experts to handle more tokens. For example, if we have 8 experts, 20 tokens, and a threshold of 0.6, as shown in Fig.~\ref{fig:Full-brownout}, we need to select 12 tokens for their corresponding experts to process, while ignoring the other 8 tokens. Therefore, we will choose E1, E3, E7, these three experts who can handle no less than 12 tokens and meet the established requirements. In this example, we processed 60\% of the tokens only accessing 37.5\% of the experts.
% \begin{figure}[htbp]
% \centerline{\includegraphics[width=0.45\textwidth]{figs/zero_brownout.png}}
% \caption{Two brownout strategies illustration with 8 experts (E0-E7) processing 20 tokens individually.}
% \label{fig:zero_brownout}
% \end{figure}

\begin{figure}[tbp]
  \centering
  \begin{minipage}[b]{0.24\textwidth}
    \centering
    \includegraphics[width=\textwidth]{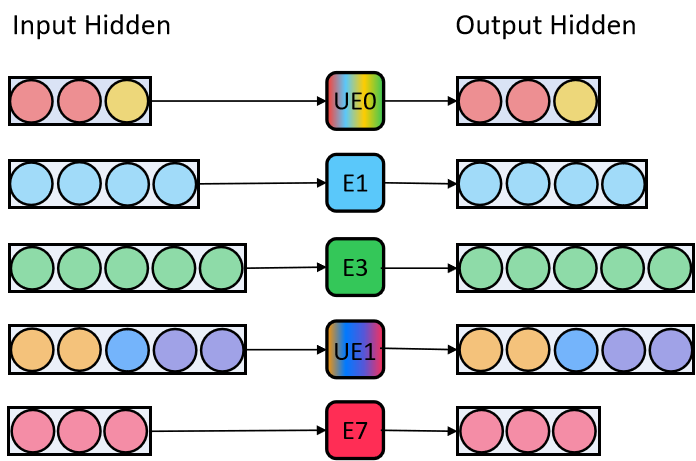}
    \subcaption{Common case.}
    \label{fig:normal_case}
  \end{minipage} \hfill
  \begin{minipage}[b]{0.24\textwidth}
    \centering
    \includegraphics[width=\textwidth]{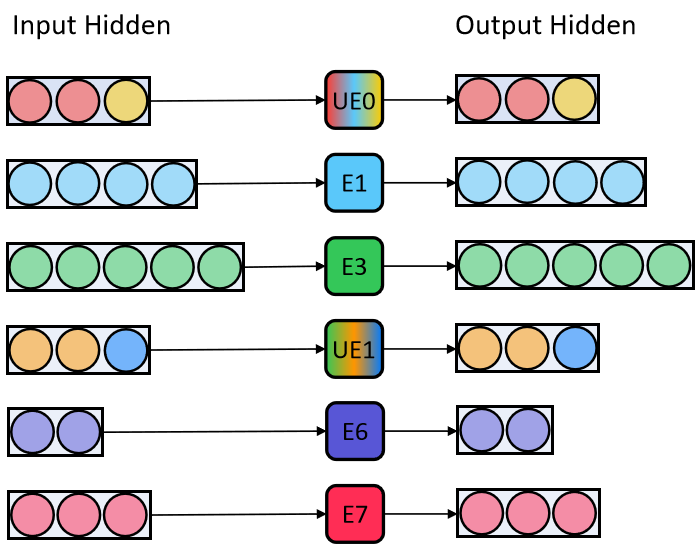}
    \subcaption{Special case.}
    \label{fig:sepcial_case}
  \end{minipage}
  \caption{Two type cases of partial-brownout strategy.}
    \label{fig:half-brownout strategies}
    \vspace{-1em}
\end{figure}

\textbf{Partial-Brownout.} This is an improved approach based on full-brownout. Unlike full-brownout, which ignores the processing of some tokens, the idea of partial-brownout is delegating these tokens that were originally ignored in the full-brownout to their corresponding united experts, thereby reducing inference latency while ensuring a certain level of inference accuracy. Suppose each transformer layer has $m$ experts, in addition to the parameter $threshold$, we also need the size of the expert group $k$, as well as $\lceil \frac{m}{k} \rceil $ united experts trained in advance through knowledge distillation techniques. Based on the example mentioned in full-brownout, we assume each transformer layer has 8 experts, E0-E7, with each expert group consisting of 4 experts (i.e., $k=4$). The number of tokens is 20, and the number of tokens each expert needs to process is 2, 4, 1, 5, 2, 1, 2 and 3 respectively. With a threshold of 0.6, as in the example in Fig.~\ref{fig:normal_case}, 12 tokens need to be processed by the original experts. Thus, we let E1, E3, and E7 handle the corresponding tokens, while E0, E2, E4, E5, and E6 delegate their tokens to the corresponding united experts instead of ignoring them. If each group contains 4 original experts, there are 2 expert groups. UE0 needs to process the tokens of E0 and E2 (a total of 3 tokens), while UE1 needs to process the tokens of E4, E5 and E6 (a total of 5 tokens). This way, we only need to access 5 experts (3 original experts and 2 united experts) to process all tokens, compared to the 8 times of expert access in zero-brownout, reducing the times of expert accesses by 3, and also increasing the batch size that experts can process at one time. Since the united experts integrate the knowledge of all experts in the group, this approach not only enhances inference efficiency, but also effectively mitigates the loss of model accuracy caused by the full-brownout strategy.

\textbf{Special Case in Partial-Brownout.} In the partial-brownout strategy, a special scenario arises when a united expert is designated to process tokens that originate solely from one original expert. Under such circumstances, it becomes more accurate to allow the original expert to handle these tokens directly rather than involving the united expert. This is because the involvement of the united expert does not contribute to reducing the times of expert accesses; on the contrary, it may potentially degrade the inference accuracy. Referring to Fig.~\ref{fig:sepcial_case}, if we set $k$=3, with a total of 8 experts, they can be divided into 3 groups. UE0 is required to process tokens from E1 and E2 (a total of 3 tokens), and UE1 is tasked with handling tokens from E4 and E5. However, the tokens from E6 are not delegated to UE2 for processing but are instead processed directly by E6 itself.

\textbf{Algorithm Design.} To decide when to trigger brownout approach, we propose Algorithm~\ref{algorithm:1}, which illustrates the specific execution process of the three brownout strategies mentioned above. After passing through the gating unit of the MoE module, we will obtain the number of tokens processed by each expert and their hidden states. The number of original experts, denoted as $m$, can be obtained from the model configuration. The threshold in the brownout method $threshold$ and the parameter $use\_full\_brownout$ (whether to enable full brownout) can be obtained from the brownout configuration, which is set by the user in advance (line~\ref{line:inputs}). After passing through the MoE module, the output of these tokens is stored in the variable $outputs$ (line~\ref{line:outputs}). The input tokens are sorted and then divided into two sets. Tokens in the set $S1$ are processed by the original experts without any accuracy loss, while tokens in the set $S2$ are processed by the united experts, which may incur some accuracy loss (lines~\ref{line:group_start}–\ref{line:group_end}). Tokens in $S1$ are directly processed by the original experts (lines~\ref{line:original_start}–\ref{line:original_end}). For the experts in $S2$ they are first grouped based on $m$ and $k$. All tokens in each expert group are concatenated and processed by the united expert (lines~\ref{line:united_start}–\ref{line:united_end}), including special case (lines~\ref{line:sepcial_case_start}–\ref{line:sepcial_case_end}) and common case (lines~\ref{line:normal_case_start}–\ref{line:normal_case_end}). Clearly, when $threshold$ is equal to 1, Algorithm~\ref{algorithm:1} describes the zero brownout process. When $ 0 \leq threshold < 1 $ and $use\_full\_brownout$ is $true$, the algorithm describes the full-brownout process. When $ 0 \leq threshold < 1 $ and $use\_full\_brownout$ is $false$, it describes the partial-brownout process.

\textbf{Time Complexity Analysis.} Algorithm~\ref{algorithm:1} has $m$ original experts  and $\left\lceil \frac{m}{k} \right\rceil$ united experts. It sorts the input tokens first and divides them into two sets: $S1$ and $S2$, then processes these tokens by original experts and united experts respectively, so the total time complexity is $O(mlogm+m+m+\left\lceil \frac{m}{k} \right\rceil)$, which equals to $O(mlogm+2m+\left\lceil \frac{m}{k} \right\rceil)$.

\begin{algorithm}
\caption{BrownoutMoE Algorithm}\label{algorithm:1}
\begin{algorithmic}[1]
\State \textbf{Input:} A tuple $(id_i, cnt_i, h_i)$ representing the id of i-th original expert $id_i$, the number of tokens to be processed $cnt_i$, and the hidden state for each token $h_i$, the number of original experts $m$, ways of the united experts $k$, threshold in brownout approach $threshold$, brownout strategy $use\_full\_brownout$. \label{line:inputs}
\State \textbf{Output:} The result all tokens processed $outputs$ \label{line:outputs}
\State Initialize $outputs$
\State $A \xleftarrow{} $ {\small $[(id_0, cnt_0, h_0), (id_1, cnt_1, h_1), \dots, (id_{m-1}, cnt_{m-1}, h_{m-1})]$}\label{line:group_start}
\State $A \gets \text{sort}(A, \text{key} = \text{cnt}, \text{descending})$
\State $S \gets \text{sum}(cnt \text{ for } (_, cnt,) \in A)$
\State $T \gets S \times threshold$
\State $sum\_partial \gets 0$
\For{$i = 0$ \textbf{to} $m-1$}
    \If{$sum\_partial \geq T$ \textbf{and} ($i = 0$ \textbf{or} $sum\_partial - A[i].cnt < T$)} \State add $A[i]$ to $S1$
    \ElsIf{not $use\_full\_brownout$}
            \State add $A[i]$ to $S2$
    \EndIf
     \State $sum\_partial\ +=\ A[i].cnt$
\EndFor \label{line:group_end}
\For{each $e \in S1$} \label{line:original_start}
    \State process\_tokens($e.id, e.h, outputs$)
\EndFor \label{line:original_end}
\State $num\_groups \gets \left\lceil \frac{m}{k} \right\rceil$ \label{line:united_start}
\State $grouped\_experts \gets \text{group\_experts}(S2, num\_groups, k)$
\For{each $group \in grouped\_experts$}
    \If{$len(group) = 1$}\label{line:sepcial_case_start}
        \State process\_tokens($e.id, e.h$)\label{line:sepcial_case_end}
    \Else
        \State $concated\_tokens \gets \text{concat\_tokens}(group)$\label{line:normal_case_start}
        \State $expert\_id \gets \text{get\_united\_expert\_id(group)}$
        \State  {\small process\_tokens($expert\_id, concated\_tokens, outputs$)}\label{line:normal_case_end}
    \EndIf
\EndFor \label{line:united_end}
\State \textbf{Return} $outputs$
\end{algorithmic}
\end{algorithm}

\subsection{Architecture of BrownoutMoE}\label{sec:BrownoutMoE}
% \vspace{-1.2em}
Compared to existing MoE architectures such as DeepSeekMoE, which have been widely applied in the DeepSeek series of models, BrownoutMoE introduces united experts to assist in inference, with the aim of reducing inference latency during emergency periods and guaranteeing SLOs. Fig.~\ref{fig:brownoutMoE} illustrates the brownoutMoE architecture and we will provide a detailed introduction to the architecture in this section. $\textbf{x}_t$ denotes the t-th token input to the MoE module. Then, the output $\textbf{h}_t$ of the MoE module can be calculated as below:
\begin{figure}[htbp]
  \begin{center}
  \hspace*{0.4cm} 
  \includegraphics[width=4.0in]{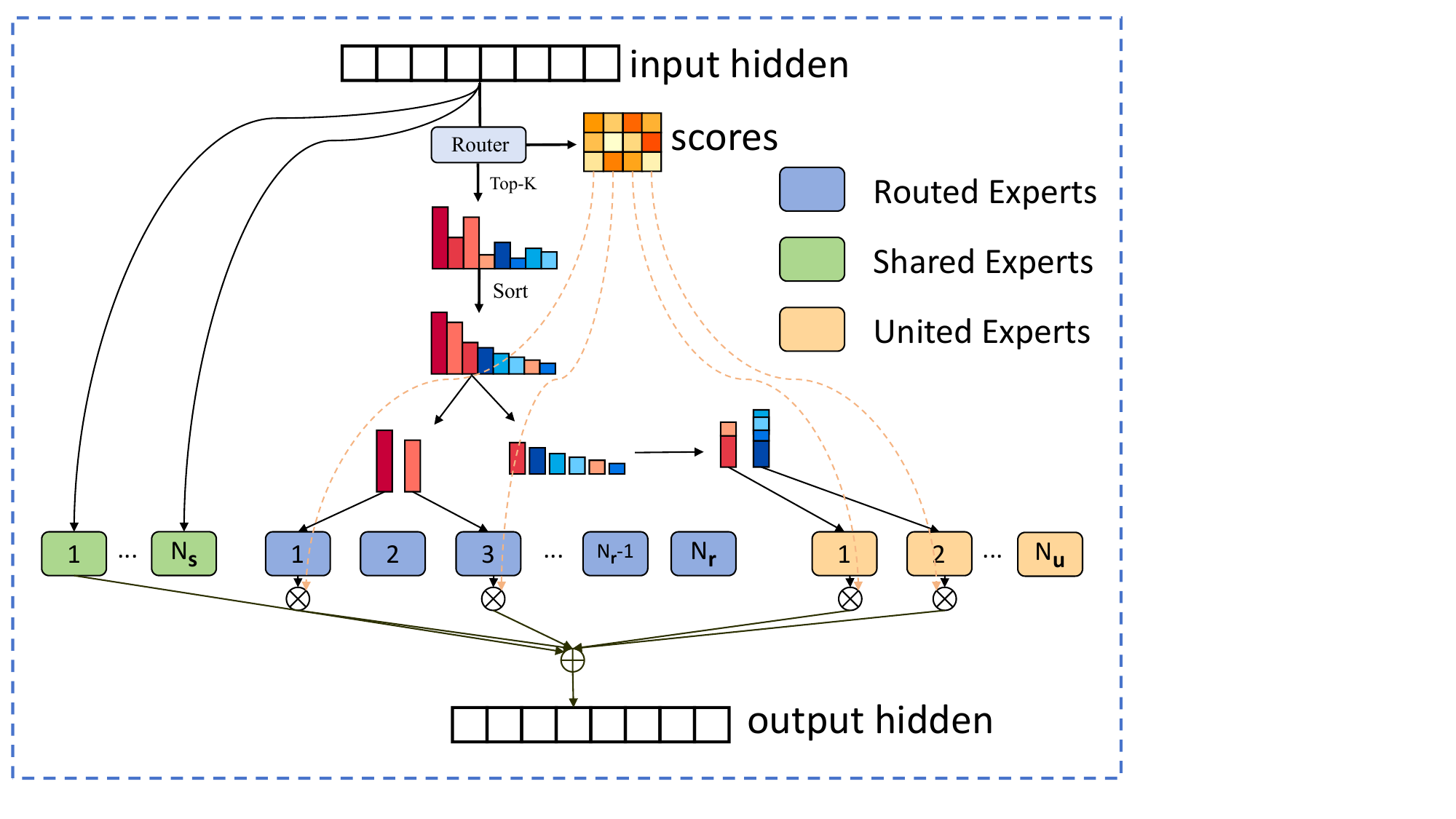}\\
\caption{Illustration of MoE with Brownout Approach. }
  \label{fig:brownoutMoE}
  \end{center}
  \vspace{-1em}
\end{figure}
\small
\begin{equation}   
% \begin{samepage}
%\scalebox{0.80}{$
\textbf{h}_t = \textbf{x}_t + \sum_{i=1}^{N_s} \text{FFN}_i^{(s)}(\textbf{x}_t) + \sum_{i=1}^{N_r} p_{i,t} \text{FFN}_i^{(r)}(\textbf{x}_t)  + \sum_{i=1}^{N_u} q_{i,t} \text{FFN}_{f(i)}^{(u)}(\textbf{x}_t)
,
%$}
\end{equation}
\normalsize 
where $N_s$, $N_r$, $N_u$ are the numbers of shared experts, routed experts (original experts), and united experts, respectively; $\text{FFN}_i^{(s)}(*)$ and $\text{FFN}_i^{(r)}(*)$ 
denote the i-th shared expert and the i-th routed expert, respectively; $\text{FFN}_{f(i)}^{(u)}(*)$ denotes the united expert corresponding to the i-th original expert. $p_{i,t}$ denotes the score between the t-th token and the i-th original expert if this token belongs to set $S1$. while $q_{i,t}$ refers to the score between the t-th token and the i-th original expert if this token belongs to set $S2$. And they can be describe as:

\begin{equation}
p_{i,t} = 
\begin{cases}
g_{i,t}, & \text{if } i \in S1, \\
0, & \text{otherwise.}
\end{cases}
,\quad
q_{i,t} = 
\begin{cases}
g_{i,t}, & \text{if } i \in S2, \\
0, & \text{otherwise.}
\end{cases}
,\quad
\end{equation}
where $S1$ and $S2$  denote the set that needs to be processed by the original experts and that needs to be processed by the united experts respectively, as shown in Algorithm~\ref{algorithm:1}. $g_{i,j}$ refers to the score between the i-th token and the i-th expert, which can be describe as:

{\small
\begin{equation}
g_{i,t} = 
\begin{cases} 
\frac{\exp(s_{i,t})}{\sum_{j \in \mathrm{TopK}} \exp(s_{j,t})}, & \text{if } i \in \mathrm{TopK}(\{s_{j,t} | 1 \leq j \leq N_r\}, K), \\
0, & \text{otherwise}.
\end{cases}
,\quad
\end{equation}
}where $TopK(*,K)$ is the set of scores consisting of the $K$ highest scores among the affinity scores calculated for the t-th token and all routed experts. $s_{i,t} $ is the affinity between t-th token and i-th expert, which is calculated by the gate function, and can be describe as: 

\begin{equation}
s_{i,t} = \textbf{x}_t^T\textbf{e}_i
,
\end{equation}
where $\textbf{e}_i$ is the centroid vector of the i-th routed expert.

\subsection{SLO-Aware Latency Control Algorithm}\label{sec:salc}
\textbf{Tradeoff Analysis.} In the brownout configuration, the parameters $k$ and $threshold$ determine the effectiveness of the brownout mechanism. A larger $k$ and a smaller $threshold$ typically yield greater improvements in inference latency. However, they also lead to increased accuracy degradation. Intuitively, a larger $k$ means that each united expert needs to integrate knowledge from more original experts, while its parameter capacity remains fixed, thus reducing the amount of knowledge it can capture from each individual expert. A smaller $threshold$ implies that more tokens are routed to united experts. Therefore, a trade-off must be made between inference latency and prediction accuracy. A common goal is to maximize the average prediction accuracy while ensuring that each token's latency meets the predefined SLO, as expressed in the following formula:

\begin{align}
\quad & \text{maximize} \   \frac{1}{n} \sum_{i=1}^{n} \text{Accuracy}^i(threshold, k), \notag \\
\quad & \text{subject to }  \text{Latency}^i_j(threshold, k) \leq \text{SLO}, \notag\\
\quad & \forall i \in \{1, 2, \dots, n\}, \forall j \in \{1, 2, \ldots, m_i\},  \label{eq:tradeoff1}
\end{align}
where $n$ is the total number of requests to be processed, $\text{Accuracy}^i(*,*)$ denotes whether the i-th request' answer is correct, $\text{Latency}^i_j(*,*)$ is latency  of the i-th request's j-th token, and $m_i$ is the i-th request's output length.  In actual inference scenarios, the parameter $k$ for brownout approach changes much less frequently than $threshold$. This is because a change in $k$ implies unloading the current united experts from the GPU memory to CPU memory or disk, and then loading the new united experts back into GPU memory. This incurs non-trivial GPU bandwidth overhead. On the other hand, the overhead introduced by changes in $threshold$ is negligible, and as we will see in the following sections, $threshold$ can be reconfigured at each iteration. Therefore, during a certain inference period, $k$ can be treated as a constant value. $\frac{1}{n} \sum_{i=1}^{n} \text{Accuracy}^i(threshold, k)$ is a monotonically increasing function with respect to $threshold$. Consequently, our target is to maximize $threshold$ while ensuring that each token's latency meets the predefined SLO.

\textbf{Algorithm Design.} Considering that the factors affecting inference latency are diverse and unpredictable, including fluctuations in the request rate, the variability in request input/output lengths, and interference from other services located on the same machine with the LLM inference service, it is challenging to determine an optimal $threshold$ in advance to satisfy Eq.~(\ref{eq:tradeoff1}). Therefore, we need to dynamically adjust $threshold$ according to real-time latency feedback during serving. Based on this insight, we design the SALC algorithm, shown as Algorithm~\ref{algorithm:2}.

\begin{algorithm}[tbp]
\caption{SLO-Aware Latency Control Algorithm}
\label{algorithm:2}
\begin{algorithmic}[1]
\State \textbf{Input:} Current brownout threshold $threshold$, requests' SLO $slo$, SLO warning line factor $warning\_factor$, recent time window $tw$,  the linear increment of threshold $increment$, the multiplicative shrinkage ratio of threshold $shrink\_ratio$.
\State \textbf{Output:} updated threshold $threshold$.
\State $warning\_line \gets slo \times warning\_factor$ \label{line:warning_line}
\State{$latency \gets $get\_recent\_P90\_latency($threshold$, $tw$)}\label{line:latency}
\If{$latency < warning\_line$}\label{line:increase_start}
    \State {$threshold \gets threshold + increment$}\label{line:increase_end}
\ElsIf{$latency > slo$}\label{line:decrease_start}
    \State{$threshold \gets threshold \times shrink\_ratio$}\label{line:decrease_end}
\EndIf
\State \textbf{Return} $threshold $
\end{algorithmic}
\end{algorithm}

The core idea of SALC is to keep latency slightly below the SLO level, as excessively low latency offers little benefit to the service and may result in a significant loss of prediction accuracy. Therefore, SALC  introduces a SLO warning line called $warning\_line$, which is a fraction of the SLO, defined with $warning\_factor$ (line~\ref{line:warning_line}). When the recent latency exceeds the SLO, $threshold$ is much reduced to lower the latency below the SLO through multiplying it by $shrink\_ratio$  (lines~\ref{line:decrease_start}-~\ref{line:decrease_end}). When the latency becomes too low (i.e., below the warning line), we increase $threshold$ linearly (lines~\ref{line:increase_start}-~\ref{line:increase_end}). This ensures that the inference latency is kept within the range between the SLO warning line and the SLO.

\textbf{Time Complexity Analysis.} Algorithm~\ref{algorithm:2} processes $n$  tokens in recent time $tw$ (at line~\ref{line:latency}). To get P90 latency of all tokens, it needs to sort them first ($O(nlogn)$), and get the value at the P90 percentile ($O(1)$). The time complexity of lines~\ref{line:warning_line},~\ref{line:increase_end} and~\ref{line:decrease_end} are $O(1)$. Therefore, the total time complexity is $O(nlogn+1\times4)$, which equals to $O(nlogn)$.

% \subsection{Rolling Update} \label{sec:rolling_update}
% During the process of the model serving, when faced with constantly changing demands, such as balancing between different precision and computational resource requirements, we may need to dynamically update the united experts without restarting the service. This demand not only requires greater flexibility but also ensures the stability and continuity of the service. Therefore, adopting a rolling update mechanism to gradually replace the united experts becomes an ideal solution. The steps for the united experts rolling update are as follows:
% \begin{itemize}
% \item \textbf{Step 1:} A subset of new experts is loaded into the GPU memory.
% \item \textbf{Step 2:} These new experts are then used to replace the corresponding old experts, and the tokens are redirected to be processed by these new experts.
% \item \textbf{Step 3:} The old experts that have been replaced are removed from the GPU memory to the CPU memory.
% \item \textbf{Step 4:} Steps 1 to 3 are repeated until all old experts have been replaced by the new experts.
% \end{itemize}

% The advantage of rolling updates is that by gradually unloading old experts and incrementally loading new experts, the system ensures it can provide stable service to users at all times. Each update involves replacing only a portion of the united experts, which helps minimize service interruptions, performance fluctuations, and resource overload, thus maintaining the efficient and stable operation of the inference service.

\section{Evaluations}
In this section, we introduce our experimental settings and evaluate the performance under a variety of workloads.

\subsection{Experimental Setup}
\textbf{Model and server configurations.} We use Qwen1.5-MoE-A2.7B-Chat~\cite{Qwen2025} for evaluation. The model has a total of 14.3B parameters, with 60 experts per transformer layer. Compared to other popular MoE models, such as Mixtral-8x7B, which has only 8 experts per layer, Qwen1.5-MoE-A2.7B-Chat contains more experts in each layer, making the brownout approach more effective for this type of model. All experiments are run on a machine equipped with 4 NVIDIA A100-PCIE-40GB GPUs (40GB  memory per GPU) and an Intel Xeon Gold 6238 processor. 

\begin{figure}[tbp]
  \centering
  \begin{minipage}[b]{0.24\textwidth}
    \centering
    \includegraphics[width=\textwidth]{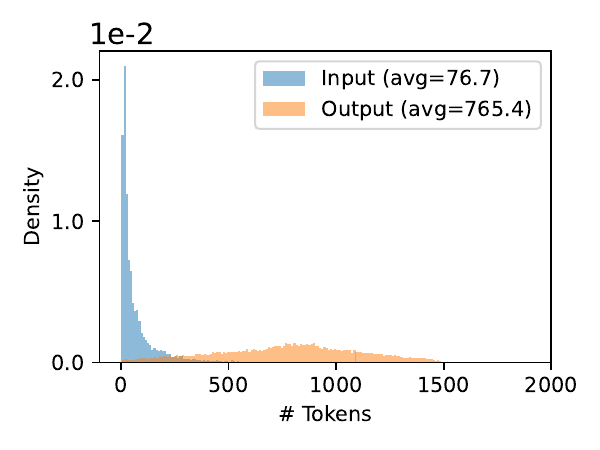}
    \subcaption{ShareGPT}
    \label{fig:sharegpt_tokens_distribution}
  \end{minipage} \hfill
  \begin{minipage}[b]{0.24\textwidth}
    \centering
    \includegraphics[width=\textwidth]{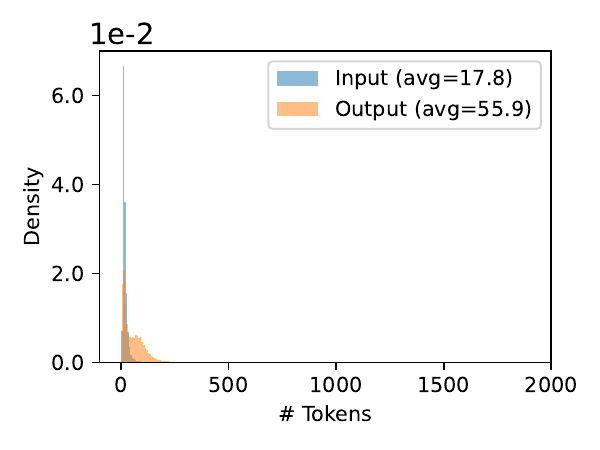}
    \subcaption{Alpaca}
    \label{fig:alpaca_tokens_distribution}
  \end{minipage}
  \caption{The input and output tokens length distributions of (a) ShareGPT datasets, (b) Alpaca datasets. }
    \label{fig:token_distributions}
\end{figure}

\textbf{Workloads.} To evaluate the prediction accuracy of the model adopting the brownout approach, We use four few-shot tasks: PIQA~\cite{Bisk2020}, COPA~\cite{roemmele2011choice}, CEVAL~\cite{huang2023ceval}, and OBQA~\cite{OpenBookQA2018}. These datasets are widely used as standard benchmarks for model accuracy evaluation in related works~\cite{liu2024deepseek,lee2024infinigen}. For fine-grained evaluation of model inference performance, we use the ShareGPT~\cite{sharegpt2025} and Alpaca~\cite{alpaca} datasets, which are also used in the evaluation of vLLM. The ShareGPT dataset is a collection of user-chatbot conversations with ChatGPT. The Alpaca dataset, is an instruction tuning dataset designed for fine-tuning or evaluating language models. Both datasets serve as typical representations of real-world LLM services. As shown in Fig.~\ref{fig:token_distributions}, the input of ShareGPT is approximately 4.3$\times$ longer than that of Alpaca, and the output of ShareGPT is approximately 13.7$\times$ longer than that of Alpaca.

\textbf{Baseline: vLLM~\cite{kwon2023efficient}.} It is a high-performance inference engine for large language models, designed for high throughput and low latency in online serving scenarios. It supports a variety of state-of-the-art optimization techniques such as ContinuousBatching, FlashAttention, and PagedAttention, all of which are also integrated into BrownoutServe. By choosing vLLM as the baseline, we can control for other optimization factors and isolate the effect of the brownout approach, allowing for a clearer evaluation of its performance benefits. We evaluate two versions of vLLM:
\begin{itemize}
\item \textbf{vLLM (native):} the official release of vLLM, which uses fused MoE\footnote{Fused MoE is suitable for scenarios with strict requirements on computational efficiency and resource consumption, while standard MoE offers more advantages in diversity, flexibility, and task adaptability.} by default. This version is used to evaluate the performance gains of the brownout approach on models that adopt fused MoE.

\item \textbf{vLLM (non-fused):} a modified version of vLLM where the fused MoE module is replaced with a standard MoE implementation. This version serves to assess the effectiveness of the brownout approach on models that do not employ fused MoE optimization.
\end{itemize}

\begin{table}[t]
\centering
\begin{tabular}{c|c|c|c|c}
\toprule
\multirow{2}{*}{\textbf{Threshold}} & \multicolumn{4}{c}{\textbf{Ways}} \\
\cline{2-5}
 & \textbf{2} & \textbf{4} & \textbf{8} & \textbf{full brownout} \\ 
\midrule
1.0 & 0.00\% & 0.00\% & 0.00\% & 0.00\% \\
0.9 & 0.06\% & 0.03\% & 1.01\% & 0.40\% \\
0.8 & 0.88\% & 0.80\% & 1.59\% & 4.51\% \\
0.7 & 1.52\% & 1.58\% & 2.29\% & 7.39\% \\
0.6 & 2.34\% & 2.38\% & 2.66\% & 9.58\% \\
0.5 & 2.81\% & 4.37\% & 3.78\% & 11.51\% \\
0.4 & 2.92\% & 3.63\% & \textbf{\textcolor{green}{4.82\%}} & 35.07\% \\
0.3 & 3.96\% & 4.43\% & 7.92\% & 59.95\% \\
0.2 & 3.30\% & \textbf{\textcolor{green}{5.18\%}} & 10.71\% & 95.08\% \\
0.1 & 4.23\% & 9.38\% & 12.60\% & 98.69\% \\
0.0 &\textbf{ \textcolor{green}{4.70\%}} & 11.53\% & 15.77\% & 100.00\% \\
\bottomrule
\end{tabular}
\caption{Average accuracy loss (\%) of Qwen1.5-MoE-A2.7B-Chat across different thresholds and ways.}
\label{tab:average_loss}
\vspace{-1.5em}
\end{table}
\begin{figure}[t]
  \centering
  \begin{minipage}[b]{0.24\textwidth}
    \centering
    \includegraphics[width=\textwidth]{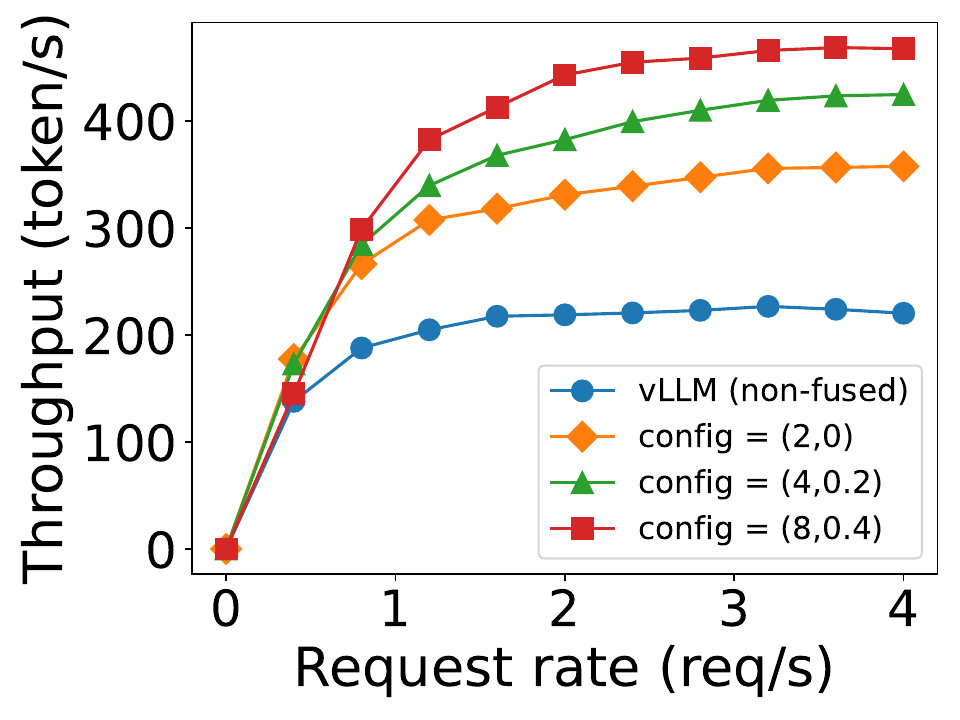}
    \subcaption{ShareGPT}
    \label{fig:unfused_throughput_sharegpt}
  \end{minipage} \hfill
  \begin{minipage}[b]{0.24\textwidth}
    \centering
    \includegraphics[width=\textwidth]{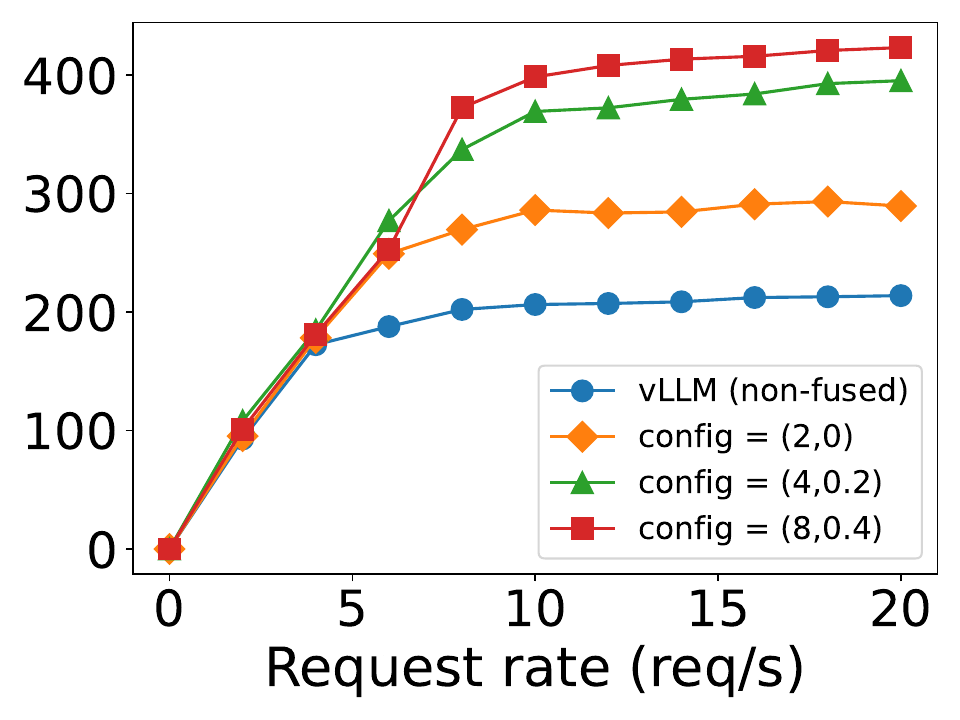}
    \subcaption{Alpaca}
    \label{fig:unfused_throughput}
  \end{minipage}
  \caption{Throughput comparison of models without fused-MoE on ShareGPT and Alpaca datasets. }
    \label{fig:unfued_throughput}
\end{figure}
\begin{figure}[t]
  \centering
  \begin{minipage}[b]{0.24\textwidth}
    \centering
    \includegraphics[width=\textwidth]{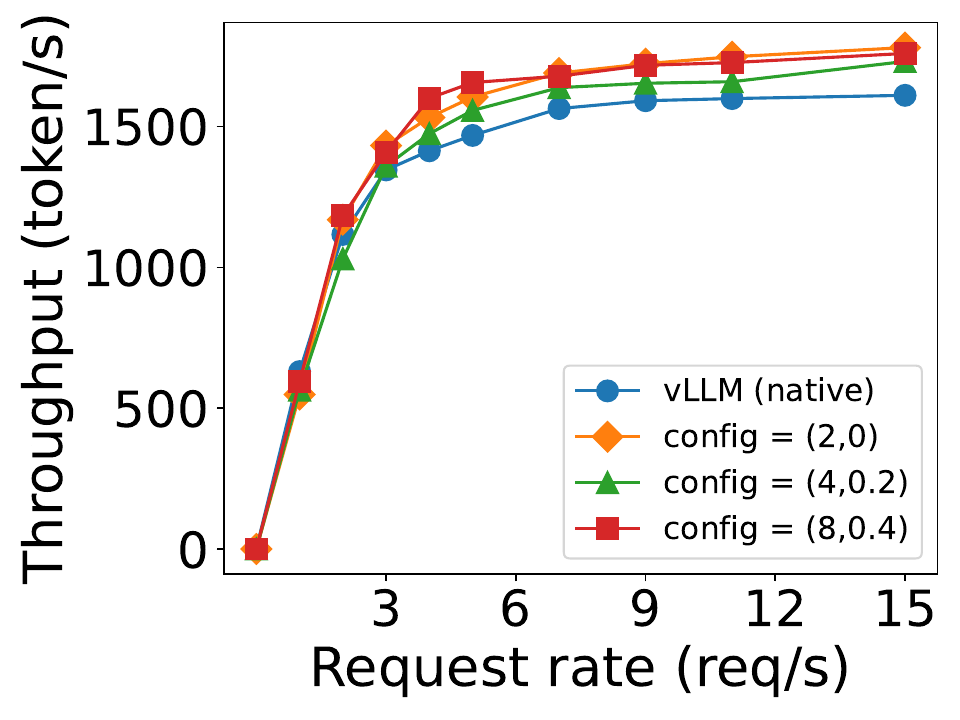}
    \subcaption{ShareGPT}
    \label{fig:fused_throughput_sharegpt}
  \end{minipage} \hfill
  \begin{minipage}[b]{0.24\textwidth}
    \centering
    \includegraphics[width=\textwidth]{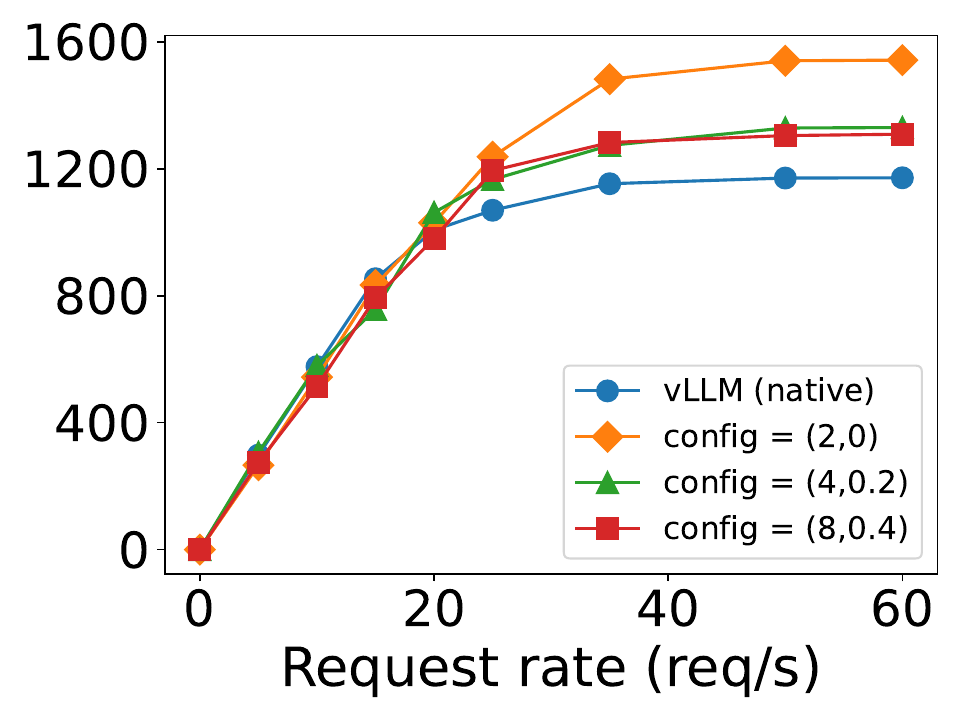}
    \subcaption{Alpaca}
    \label{fig:fused_throughput_alpaca}
  \end{minipage}
  % \vspace{-3mm}
  \caption{Throughput comparison of models with fused-MoE on ShareGPT and Alpaca datasets. }
    \label{fig:fused_throughput}
    \vspace{-1em}
\end{figure}
We mainly evaluate the following three key metrics:
\begin{itemize}
\item \textbf{Throughput:} To provide a more comprehensive and detailed evaluation of throughput , we test three different ($way$, $threshold$) configurations: (2, 0), (4, 0.2) and (8, 0.4). We compare the throughput against the baseline at various request rates, with a trace that requests continuously sent for 10 minutes.

\item \textbf{Accuracy:} We evaluate the accuracy loss introduced by using the brownout approach during inference under different $way$ and $threshold$ in brownout configurations.

\item \textbf{SLO Violations Rate:} It is defined as the proportion of individual tokens that fail to meet the predefined SLO. This is an important metric for assessing service reliability and user experience in real-world production environments.

\end{itemize}
\begin{figure*}[t]
  \begin{center}
  \includegraphics[width=7in]{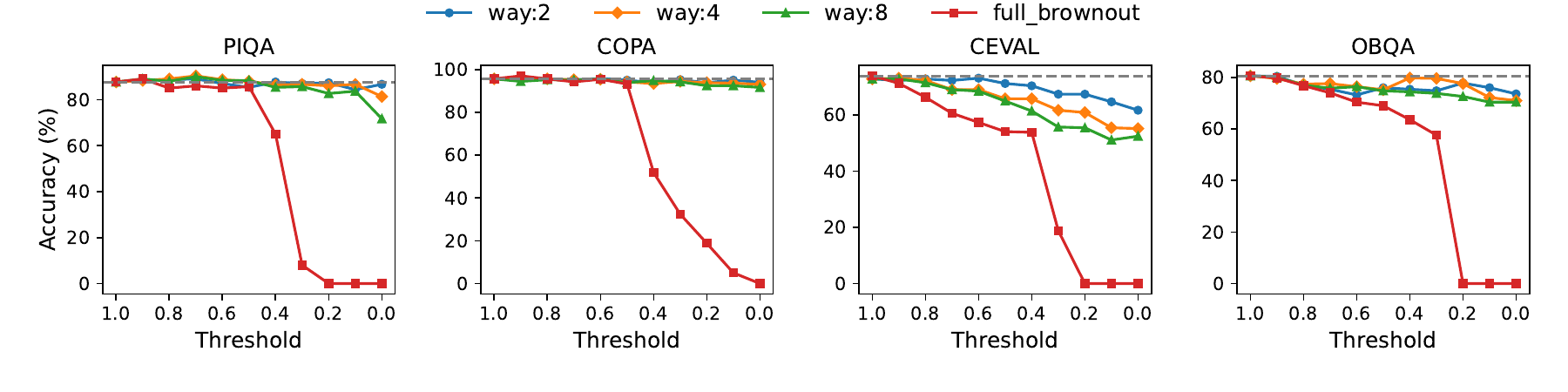}\\
\caption{Accuracy of Qwen1.5-MoE-A2.7B-Chat on 5-shot tasks.}
  \label{fig:accuracy}
  \end{center}
    \vspace{-1.5em}
\end{figure*}

\begin{figure*}[t]
  \begin{center}
  \includegraphics[width=7in]{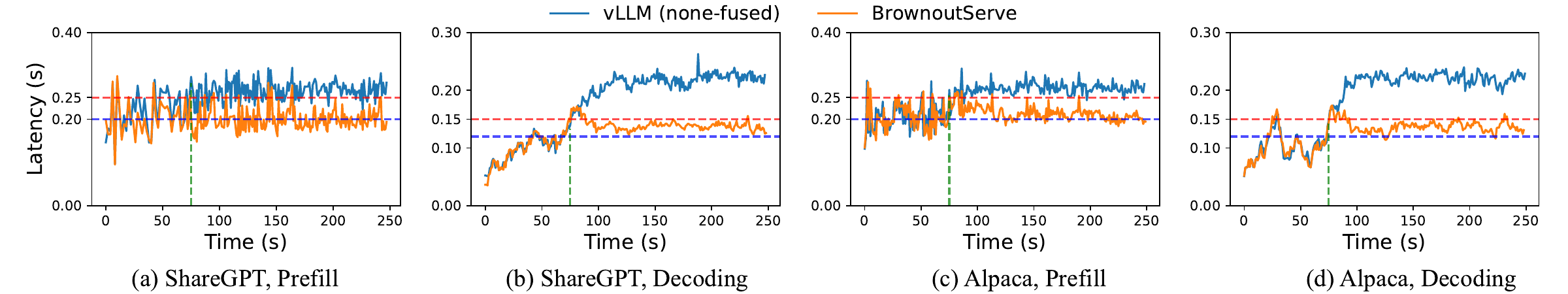}\\
\caption{P90 latency traces comparing vLLM (non-fused) and  under 8-way united experts.}
  \label{fig:latency_trace}
  \end{center}
  \vspace{-1.5em}
\end{figure*}

\subsection{Throughput}\label{throughput}
According Eq. (\ref{eq:request_modeling}), We configure clients that continuously send requests to  for 10 minutes with different request rates, where the request arrival follows a Poisson distribution. Serving throughput is defined as the ratio between the total number of tokens generated and the total time consumed. Given the practical value of the configurations (2, 0), (4, 0.2), and (8, 0.4), we evaluate BrownoutServe’s performance under these three settings. We set the maximum batch size to 64 and the maximum sequence length to 2048. As shown in Fig.~\ref{fig:unfued_throughput}  achieves a 1.58$\times$ to 2.07$\times$ throughput improvement over vLLM (non-fused) on the ShareGPT dataset, and a 1.37$\times$ to 2.06$\times$ improvement on the Alpaca dataset. When equipped with the fused MoE module, as shown in Fig.~\ref{fig:fused_throughput}  outperforms vLLM (native) by 1.07$\times$ to 1.10$\times$ on ShareGPT dataset and 1.12$\times$ to 1.32$\times$ on Alpaca dataset. Compared to the significant improvement over vLLM (non-fused), the gains over vLLM (native) are more modest, primarily because the fused MoE operator is already highly optimized. According to Eq.~(\ref{eq:speedup}), there is limited room for further performance enhancement within the fused MoE, as it is already near optimal.

\subsection{Accuracy}
Fig.~\ref{fig:accuracy} shows the accuracy variation on four datasets (PIQA, COPA, CEVAL, OBQA) under different brownout configurations for 5-shot tasks. We designed four brownout united expert modes: 2-way, 4-way, 8-way, and full brownout. For each mode, we varied the threshold from 1 to 0 (with a step size of 0.1), resulting in 44 different brownout configurations (4 modes × 11 threshold values) in total. When the threshold is set to 1, the brownout approach actually adopts the zero-brownout strategy, which serves as the baseline for self-comparison of 's accuracy metric (indicated by the gray dashed line in the figure). As seen in the figure, on the CEVAL and OBQA datasets, the accuracy decreases more significantly as the threshold decreases, compared to the PIQA and COPA datasets. This may be because the problems in these two datasets are more difficult for the model to comprehend, making the model more sensitive to the changes in the brownout strategy when handling these more complex problems. Table~\ref{tab:average_loss} presents the average accuracy loss of model Qwen1.5-MoE-A2.7B-Chat across 44 different brownout configurations evaluated on four datasets. We specifically highlight three representative settings: (2, 0), (4, 0.2), and (8, 0.4), which result in average accuracy losses of 4.70\%, 5.18\%, and 4.82\%, respectively, around the 5\% level. For short-lived bursty workloads, tolerating an accuracy drop of approximately 5\% is a practical trade-off to meet SLO attainment. Therefore, these configurations strike a practical balance between computational savings and model performance, making them valuable for deployment.

\subsection{SLO Violations Rate}
We design a 250s request stream to evaluate traffic burst commonly encountered in real-world applications. During this process, we observe the robustness of  under sudden request pressure and validate the effectiveness of the SALC algorithm by evaluating its improvement over baseline methods in terms of SLO violations rate. We used the ShareGPT and Alpaca datasets as request sources, with an initial request rate of 0.5 requests per second (RPS) for ShareGPT and 1 RPS for Alpaca, and doubled the request rate at the 75s mark to create a bursty scenario. There we use the identical settings of batch size and sequence length in~\ref{throughput}.

\textbf{Latency Trace Analysis.} Fig.~\ref{fig:latency_trace} illustrates a detailed record of the prefill and decoding latency (P90 latency) over a 250s request trace. The red dashed lines represent the SLO (0.25s for prefill and 0.15s for decoding), while the blue dashed lines is the SLO warning lines (0.20s for prefill and 0.12s for decoding, with a warning factor of 0.8). Both systems initially met the SLO requirements during the baseline phase ($t<$ 75s). Upon the 2× load surge at $t$=75s, vLLM (non-fused) exhibited significant latency degradation, peaking at 0.31s (24\% over SLO) for prefill and 0.24s (60\% over SLO) for decoding. Regulated by SALC ($shrink\_ratio$, in Algorithm~\ref{algorithm:2} is set As 0.8 and $increment$ as 0.1), maintained stable performance within the warning-SLO buffer zone (prefill: 0.20-0.25s; decoding: 0.12-0.15s). In particular,  demonstrated a reduction 67\% in latency oscillations ($\sigma$=0.008s vs. 0.024s).

\begin{figure}[t]
  \begin{center}
  \includegraphics[width=3.5in]{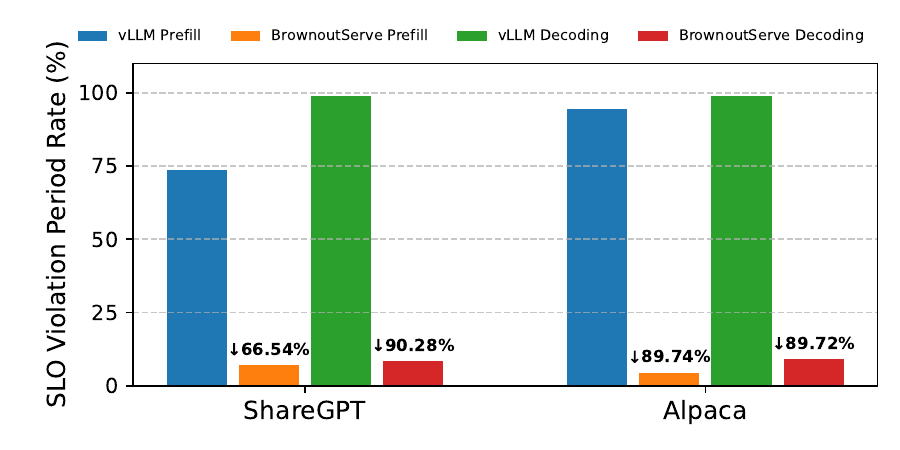}\\
\caption{SLO violations rate comparison between vLLM and  across ShareGPT and Alpaca datasets.}
  \label{fig:svpr}
  \end{center}
    \vspace{-1em}
\end{figure}

\textbf{SLO Violation Rate Analysis.} We evaluate the SLO violations rate of vLLM under bursty traffic conditions. As shown in Fig.~\ref{fig:svpr}, on the ShareGPT dataset, vLLM exhibits SLO violations rate of 73.68\% and 98.85\% during the prefill and decoding stages, respectively, while  maintaining significantly lower rates of 7.14\% and 8.57\%, corresponding to reductions of 66.54\% and 90.28\%. On the Alpaca dataset, vLLM’s SLO violations rate reaches 94.29\% and 98.86\%, while BrownoutServe’s rates are only 4.55\% and 9.14\%, with reductions of 89.74\% and 89.72\%, respectively. These results demonstrate that  BrownoutServe exhibits stronger robustness under bursty loads and, with the help of the SALC algorithm, effectively reduces the SLO violations ratio.

\begin{figure}[tbp]
  \centering
  \begin{minipage}[b]{0.24\textwidth}
    \centering
    \includegraphics[width=\textwidth]{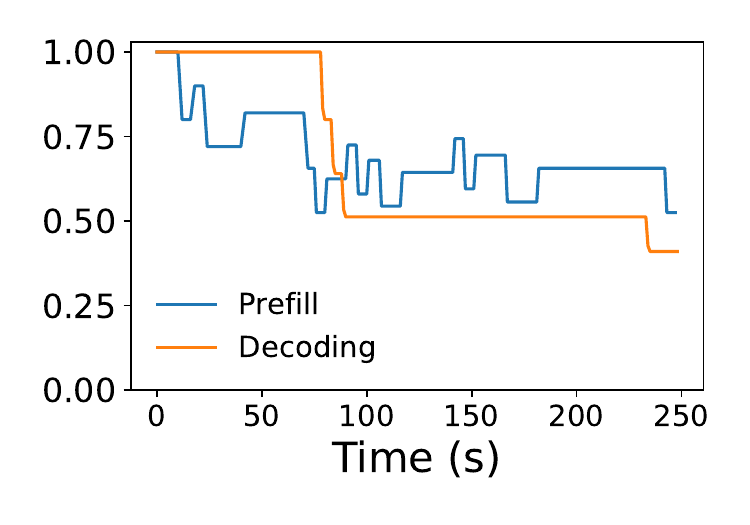}
    \subcaption{ShareGPT}
    \label{fig:sharegpt_threshold}
  \end{minipage} \hfill
  \begin{minipage}[b]{0.24\textwidth}
    \centering
    \includegraphics[width=\textwidth]{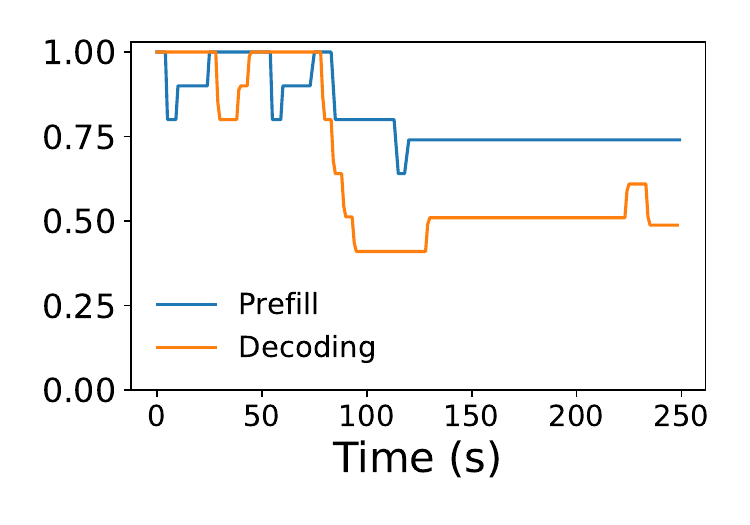}
    \subcaption{Alpaca}
    \label{fig:alpaca_threshold}
  \end{minipage}
  \caption{Threshold variations in BrownoutServe. }
    \label{fig:threshold_variation}
    \vspace{-1em} 
\end{figure}

\textbf{Threshold Variations Analysis.} We also record the threshold variations of  in this trace. As shown in Fig.~\ref{fig:threshold_variation}. During requests burst ($t>$ 75s), results show that on the ShareGPT dataset, the prefill phase achieved an average threshold of 0.635 (with $\leq$2.66\% accuracy loss at way=8, known from Table~\ref{tab:average_loss}) while the decoding phase averaged 0.524 ($\leq$3.78\% accuracy loss). For the Alpaca dataset, the prefill threshold rose to 0.760 ($\leq$2.29\% accuracy loss) with decoding stabilizing at 0.514 ($\leq$3.78\% accuracy loss). These findings demonstrate that the brownout approach effectively mitigates resource overload during traffic bursts through dynamic threshold adaptation while meeting SLO.

To summarize, BrownoutServe achieves up to 2.07$\times$ higher throughput and reduced SLO violations by 90.28\% under burst workloads compared to vLLM. This is accomplished by introducing united experts and the dynamic brownout mechanism. And using these techniques, this framework alleviates the limitations of static MoE inference systems and improves serving scalability under bursty requests.

\section{Conclusions And Future Work}
In this paper, we present BrownoutServe, an innovative serving framework for MoE-based LLMs that incorporates the brownout approach to optimize inference efficiency while maintaining service reliability under bursty workloads. Through the introduction of the united expert models and the brownout mechanism, BrownoutServe effectively reduces inference latency and enhances throughput by dynamically adjusting the allocation of computational resources based on real-time workloads. BrownoutServe's robustness and efficiency have been validated through extensive experiments under various datasets and traffic conditions. As future work, we would like to  extend our framework to multiple MoE-based LLMs, such as Mixtral and DeepSeek. Additionally, we will integrate a prefill-decoding separation architecture to further optimize SLO. 

\section*{Software Availability}
The codes of BrownoutServe have been open-sourced to \url{https://github.com/beyondHJM/BrownoutServe} for research usage.
\bibliographystyle{IEEEtran}
\bibliography{IEEEabrv,Bibliography}
\begin{IEEEbiography}[{\includegraphics[width=1in,height=1.15in,clip]{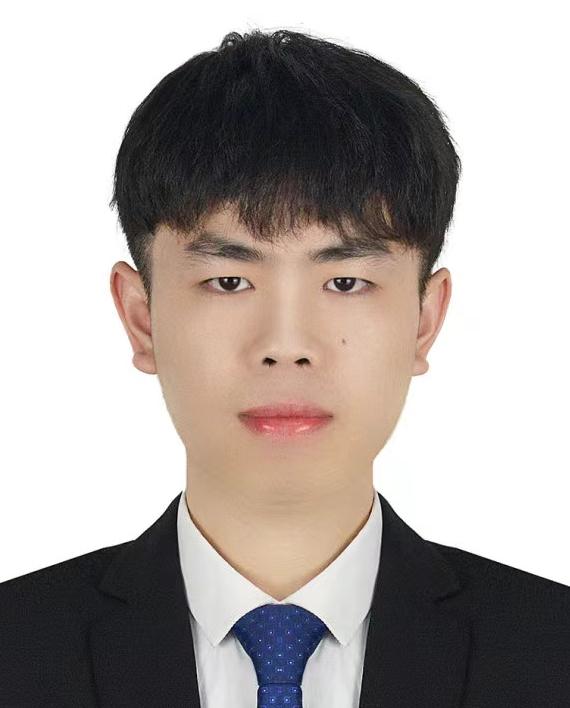}}]{Jianmin Hu}  
is currently a joint master student at Shenzhen Institutes of Advanced Technology, Chinese Academy of Sciences and Southern University of Science and Technology. He obtained a Bachelor's degree in Software Engineering from Harbin Institute of Technology in 2023. His current research direction is cloud native and system for large language model.
\end{IEEEbiography}

\begin{IEEEbiography}[{\includegraphics[width=1in,height=1.25in,clip]{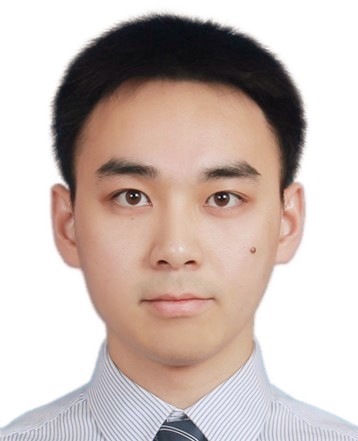}}]{Minxian Xu} (Senior Member, IEEE) is currently an Associate Professor at Shenzhen Institutes of Advanced Technology, Chinese Academy of Sciences. He received the BSc degree in 2012 and the MSc degree in 2015, both in software engineering from University of Electronic Science and Technology of China. He obtained his Ph.D. degree from the University of Melbourne in 2019. His research interests include resource scheduling and optimization in cloud computing. He has co-authored 70+ peer-reviewed papers published in prominent international journals and conferences, such as ACM CSUR, IEEE TSC, IEEE TMC, ACM TOIT, ACM TAAS and ICSOC. He was awarded the 2023 IEEE TCSC Award for Excellence (Early Career Award). He is among the world's top 2\% scientist by Stanford University. More information can be found at: minxianxu.info.
\end{IEEEbiography}

 \begin{IEEEbiography}[{\includegraphics[width=1.0in,trim=0in 0in 0in 0in,clip,keepaspectratio]{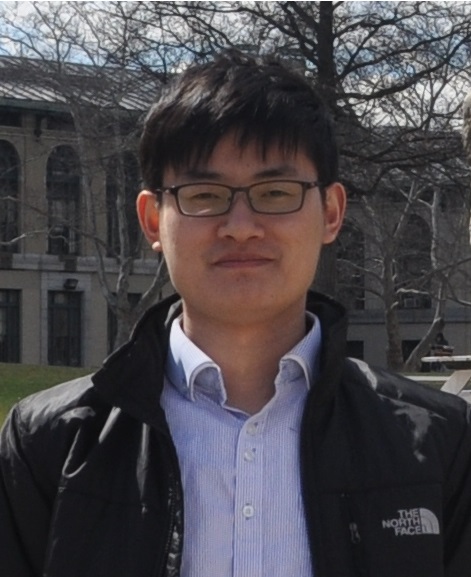}}]{Kejiang Ye} is currently a Professor and the Director of the Research Center for Cloud Computing, Shenzhen Institute of Advanced Technology, Chinese Academy of Sciences. He received his B.S. and Ph.D degrees both from Zhejiang University and was a Post Doctoral Research Associate at Carnegie Mellon University (CMU). His research interests include Digital Technology and Systems (e.g., Cloud Computing, Big Data and Industrial Internet). He is a Senior Member of IEEE and a Distinguished Member of China Computer Federation (CCF).
\end{IEEEbiography}

 \begin{IEEEbiography}[{\includegraphics[width=1.0in,height=1.25in,clip]{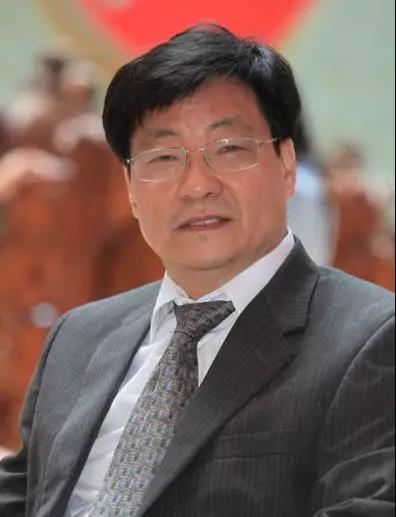}}]{Chengzhong Xu} (Fellow, IEEE) received the PhD degree in computer science and engineering from the University of Hong Kong, in 1993. He is the dean with the Faculty of Science and Technology and the interim director with the Institute of Collaborative Innovation, University of Macau. Dr. Xu’s research focuses on parallel and distributed computing, with an emphasis on resource management for performance, reliability, availability, power efficiency, and security. His work spans servers and cloud datacenters,  wireless embedded devices and edge AI systems , with applications in smart city and autonomous driving. He has authored two research monographs and more than 600 papers, which have garnered more than 22K citations with an H-index of 77, and has been cited in over 300 international patents (including 230 USA patents as of year 2024, per SciVal). Dr. Xu is a Chief Scientist of Key Project on Smart City of MOST, China and the principal investigator of the Key Project on Autonomous Driving of FDCT, Macau SAR. 
\end{IEEEbiography}
%\begin{IEEEbiographynophoto}{Ignacio Ramos}
%(S'12) received the B.S. degree in electrical engineering from the University of Illinois at Chicago in 2009, and is currently working toward the Ph.D. degree at the University of Colorado at Boulder. From 2009 to 2011, he was with the Power and Electronic Systems Department at Raytheon IDS, Sudbury, MA. His research interests include high-efficiency microwave power amplifiers, microwave DC/DC converters, radar systems, and wireless power transmission.

%% insert where needed to balance the two columns on the last page with
%%\newpage

% ==== SWITCH OFF the BIO for submission
% ==== SWITCH OFF the BIO for submission


\vfill


\end{document}